\documentclass[aps,amsmath,amssymb, notitlepage,nofootinbib,
twocolumn
]{revtex4-1}

\usepackage{graphicx}
\usepackage{color}
\usepackage{bbold}

\newcommand{\zb}{{\vec Z}}

\newcommand{\bit}{\begin{itemize}}
\newcommand{\eit}{\end{itemize}}

\newcommand{\f}{\frac}
\renewcommand{\>}{\right\rangle}
\newcommand{\<}{\left\langle}
\newcommand{\ba}{\begin{align}}
\newcommand{\ea}{\end{align}}
\newcommand{\be}{\begin{equation}}
\newcommand{\ee}{\end{equation}}
\newcommand{\bi}{\begin{itemize}}
\newcommand{\ei}{\end{itemize}}
\newcommand{\lf}{\left(}
\newcommand{\ri}{\right)}
\newcommand{\dd}{\mathrm{d}}
\newcommand{\OO}{\mathcal{O}}
\newcommand{\Or}{\mathrm{O}}
\newcommand{\SU}{\mathrm{SU}}

\newcommand{\LL}{\mathcal{L}}

\newcommand{\Tr}{\operatorname{Tr}}
\newcommand{\tr}{\operatorname{tr}}

\newcommand{\cp}{\mathrm{CP}}
\newcommand{\rp}{\mathrm{RP}}

\newcommand{\attr}{\mathcal{A}}
\newcommand{\stra}{\mathcal{S}}
\newcommand{\cros}{\mathcal{C}}

\graphicspath{{PolymerFigs/}}

\begin{document}

\newcommand{\bra}[1]{\< #1 \right|}
\newcommand{\ket}[1]{\left| #1 \>}

\title{The Generic Critical Behaviour for 2D Polymer Collapse}
\author{Adam Nahum}
\affiliation{Department of Physics, Massachusetts Institute of Technology, Cambridge, Massachusetts 02139, USA}
\date{\today}

\begin{abstract}
\noindent
The nature of the $\theta$ point  for a polymer in two dimensions has long been debated, with a variety of candidates put forward for the critical exponents. This  includes those derived by Duplantier and Saleur (DS) for an exactly solvable model. We use a representation of the problem via the $\cp^{N-1}$ $\sigma$---model in the limit $N \rightarrow 1$ to determine the stability of this critical point. First we prove that the DS critical exponents are robust, so long as the polymer does not cross itself: they can arise in a generic lattice model, and do not require fine tuning. This resolves a longstanding theoretical question. However there is an apparent paradox: two different lattice models, apparently both in the DS universality class, show different numbers of relevant perturbations, apparently leading to contradictory conclusions about the stability of the DS exponents. We explain this in terms of subtle differences between the two models, one of which is fine-tuned (and not strictly in the DS universality class). Next, we allow the polymer to cross itself, as appropriate e.g. to the quasi--2D case. This introduces an additional independent relevant perturbation, so we do not expect the DS exponents to apply. The exponents in the case with crossings will be those of the generic tricritical $\mathrm{O}(n)$ model at $n=0$, and different to the case without crossings. We also discuss interesting features of the operator content of the $\cp^{N-1}$ model. Simple geometrical arguments show that  two operators in this field theory, with very different symmetry properties, have the same scaling dimension for any value of $N$ (equivalently, any value of the loop fugacity). Also we argue that for any value of $N$  the $\cp^{N-1}$ model has a marginal parity-odd operator which is related to the winding angle.
\end{abstract}

\maketitle

\section{Introduction}

One of the most elegant ideas in polymer physics is de Gennes' mapping between long polymer chains and the $\mathrm{O}(n)$ field theory in the limit $n\rightarrow 0$ \cite{replica for loops}. The large-scale geometry of a chain in a good solvent, or a lattice self-avoiding walk, is described by the critical $\Or(n)$ model. If the solvent quality is reduced, the monomers effectively attract each other, and eventually the polymer collapses into a compact object via a phase transition known as the $\theta$-point. In de Gennes' correspondence the $\theta$ point maps to the \textit{tricritical} $\Or(n)$ model \cite{de gennes tricritical}. This has upper critical dimension three, so in three dimensions the $\theta$-point polymer is ideal (up to logarithmic corrections). The nature of the $\theta$ point in \emph{two} dimensions is much more interesting and, surprisingly, not fully understood.

\begin{figure}[t]
\includegraphics[width=0.95\linewidth]{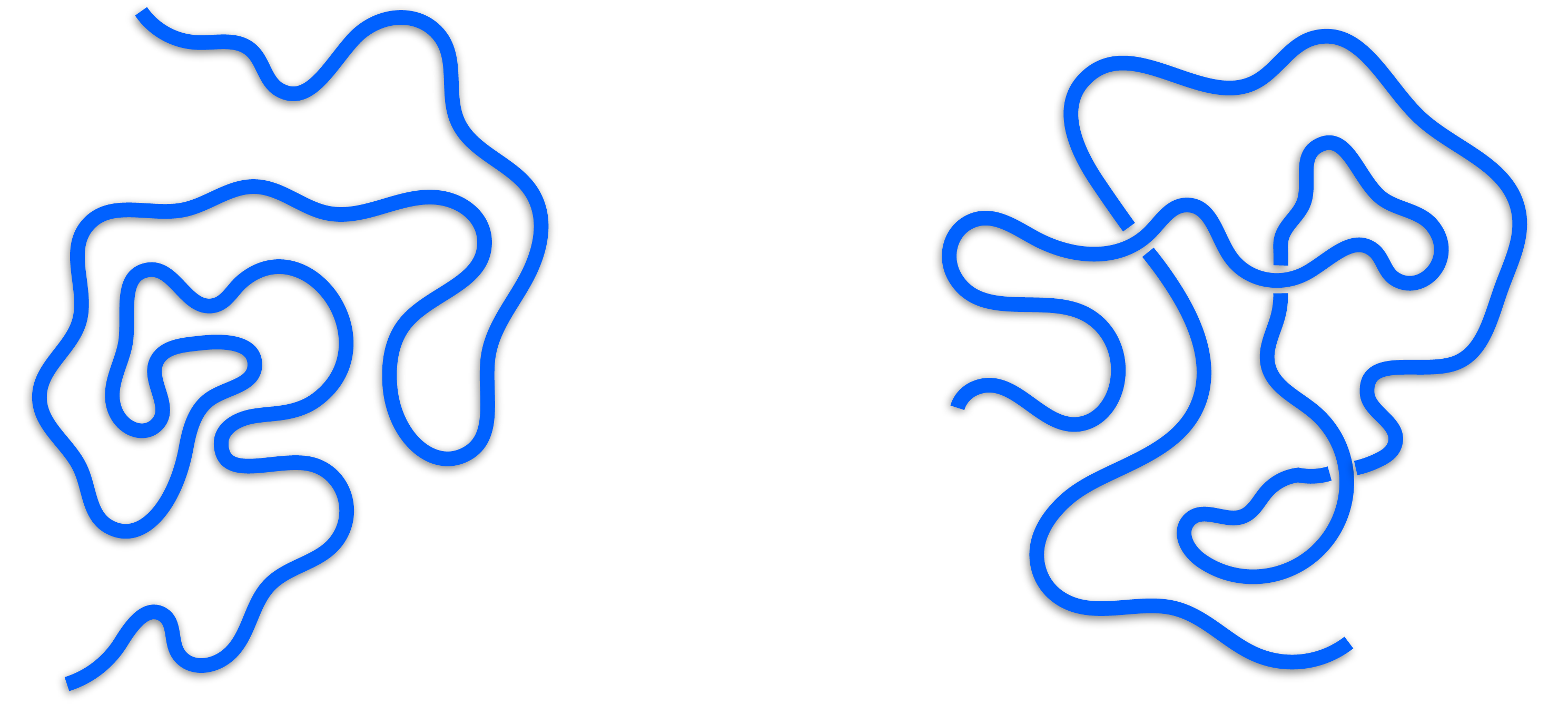}
 \caption{In 2D, a basic topological distinction is between models in which the polymer chain cannot cross itself (left) and those in which it can (right).}
 \label{withwithoutcrossings}
\end{figure}

In two dimensions we must distinguish two kinds of model according to whether or not we allow the polymer to cross itself (Fig.~\ref{withwithoutcrossings}). Most of the theoretical and numerical work has focussed on models without crossings: we discuss these first.  A key development was the derivation by Duplantier and Saleur of exact critical exponents for a particular honeycomb lattice model, in which polymer conformations have a relationship with percolation cluster boundaries \cite{duplantier saleur polymer, coniglio jan et ap}. Let us call the corresponding renormalisation group (RG) fixed point the DS fixed point. The fact that the honeycomb lattice model is only solvable at a fine-tuned point (where the correspondence with percolation holds) led to debate about whether the DS exponents captured the \textit{generic} critical behaviour at the $\theta$ point, even for non-crossing polymers. For example, Bl\"ote and Nienhuis proposed another solvable model for the $\theta$ point \cite{blote nienhuis} (which has recently attracted new interest \cite{vernier new look}), with different exponents, and argued that it should be more stable in the RG sense than the model solved by DS. On the other hand, numerical results  seem to indicate that the DS exponents are robust against changes of the model \cite{duplantier saleur numerics, seno stella,  prellberg owczarek 94, grassberger hegger}; see in particular Ref.~\cite{caracciolo}. Further complicating the issue, models are known which initially appeared to behave similarly to the DS polymer, but later turned out to show different universal behaviour with anomalously large finite size effects \cite{loops with crossings, Lyklema, Meirovitch, Owczarek and Prellberg collapse, foster universality}.

The question of what the generic universal behaviour is for the the collapse transition  has remained unresolved until now. In this paper we address it  using a representation of the DS universality class via a sigma model with $\mathrm{SU}(N)$ symmetry \cite{read saleur exact spectra} in the limit $N\rightarrow 1$. We show that the the DS exponents \textit{are} robust for \textit{non-crossing} polymers. The critical exponents of the original honeycomb lattice model \cite{duplantier saleur polymer, coniglio jan et ap} can arise in a generic model, without the need for fine-tuning. 

At the same time, there is an apparent paradox which we must resolve. At first sight one reaches contradictory conclusions about the stability of the DS point by analysing different popular models which share the same field theory description, and which at first sight are in the same universality class. We explain why this naive symmetry analysis gives misleading results. We connect this with the  fact that one of the models suffers from fine-tuning related to an Ising-like order parameter defined in Ref.~\cite{blote batchelor nienhuis 98}.

To obtain the above we classify the allowed perturbations of simple models for the $\theta$ point which show the DS exponents, making use of mappings to concrete lattice field theories \cite{cpn loops short, cpn loops long, loops with crossings, vortex lines, springer thesis}. The lattice field theories for these models have $\mathrm{SU}(N)$ symmetry. This  symmetry is `enhanced' compared to more generic polymer models: this  is a manifestation of fine-tuning of the Boltzmann weights for the polymer. Any generic perturbation to the polymer's interactions breaks the symmetry to a subgroup. However that does not in itself imply that the DS \emph{fixed point} is fine tuned. The $\SU(N)$ symmetry may be restored in the infra-red even when it is broken microscopically. We argue that this symmetry enhancement under RG is what happens for generic models in the DS universality class. The question of the robustness of the DS exponents is therefore related to the number of \emph{relevant} symmetry-breaking perturbations. (For polymers, the potential complication is that a given model may be mappable to lattice field theory in multiple ways, and an ill-chosen mapping may conceal the full symmetry.)

A generic description of the $\theta$ point should have \textit{two} relevant perturbations. Although for the polymer we need only tune one parameter to reach the $\theta$ point, the field theory is automatically tuned to criticality by taking the length of the polymer to be large. We show that polymer models that are truly in the DS universality class indeed have two relevant perturbations (when crossings are not allowed), and the fine-tuned model mentioned above has three. In order to show that no other relevant or marginal perturbations can play a role, we are also led to analyse some novel features of the relevant sigma model, the $\cp^{N-1}$ model.

A physical  polymer system in 2D or quasi-2D may allow for crossings, where one part of the polymer chain lies on top of another part (Fig.~\ref{withwithoutcrossings}). These may have an important effect at large lengthscales even if energetically disfavoured at small scales. Crossings are known to lead to new universal behaviour in completely-packed 2D loop models \cite{jacobsen read saleur, martins nienhuis rietman}. Here too, crossings may be shown to destablize the DS fixed point. Therefore in the case with crossings we do not expect the DS exponents to apply. 

Further, we argue that the tricritical $\Or(n)$ behaviour expected by de Gennes will \textit{only} be seen when crossings are allowed (i.e., the DS universality class  should not be identified with the generic tricritical $\mathrm{O}(n)$ model, contrary to what is often assumed). A subtlety here is that at first sight certain of the models we discuss have $\Or(n)$ symmetry even when the polymer cannot cross itself. However we point out that a higher  symmetry  is revealed in these models by mapping them to field theory in a different way.

The best studied model with crossings is the collapse point of the interacting self-avoiding trail  \cite{Lyklema, Meirovitch, Owczarek and Prellberg collapse, foster universality, loops with crossings, jacobsen read saleur, martins nienhuis rietman}. This model is in many ways analogous to the honeycomb lattice model solved by DS. It also has enhanced symmetry --- in this case $\mathrm{SO}(N\rightarrow 1)$, which should be regarded as larger than the $\mathrm{O}(n\rightarrow 0)$ of a generic model. Unfortunately, this critical point turns out to be infinitely fine-tuned \cite{loops with crossings}, so certainly not the generic $\theta$ point in the case with crossings! (Unlike the case with $\SU(N)$ symmetry, here there are an infinite number of relevant symmetry-breaking perturbations.)  We are not aware of any exact results for more generic models with crossings (see Ref.~\cite{bedini more generic trails} for a numerical study), and this is  an interesting subject for future research. It was suggested by de Gennes,  on the basis of the smallness of the coefficients in the $3-\epsilon$ expansion, that the  tricritical $O(n)$ exponents may be close to mean field values even in 2D \cite{de gennes tricritical}.

The field theory which is central to our analysis is the $\cp^{N-1}$ nonlinear sigma model with a $\Theta$ term at $\Theta=\pi$. In quantum condensed matter this theory is familiar from the Heisenberg spin-1/2 chain and its $\SU(N)$ generalisations \cite{Affleck SU(n) chains, fradkin book}. Its relationship with 2D loop models for loops with fugacity $N$ has been discussed extensively \cite{read saleur exact spectra, candu et al, affleck loops}. Here we are interested in the limit $N\rightarrow 1$, which \textit{a priori} describes a soup of many loops rather than  a single polymer. However, a well-known trick \cite{blote nienhuis, Cardy n+n' trick, duplantier saleur polymer, coniglio jan et ap} is to isolate a single `marked' loop and integrate out (i.e. ignore!) all the others. At $N=1$ the marked loop turns out to be governed by a \emph{local} Boltzmann weight, as appropriate to a polymer. To study generic interactions for the polymer, we must change the interactions for this marked loop without modifying the weights for the soup of background loops. This corresponds to introducing various anisotropies in the sigma model. This strategy was pursued for the $\rp^{N-1}$ sigma model describing the interacting self-avoiding trail in Refs.~\cite{loops with crossings, springer thesis}. In that case the effect of the perturbations is simpler to analyse (because the governing fixed point is  free  \cite{jacobsen read saleur}), but the logic is the same.

The analysis will lead us to examine the operator content of the $\cp^{N-1}$ sigma model. We find some features that are surprising from the point of view of field theory but transparent from the loop-gas perspective. For example, a simple geometrical argument shows that two operators in the field theory with very different properties under spatial and $\SU(N)$ symmetries (and different numbers of spatial derivatives) are forced to have the same scaling dimension for any $N$. This is related to the results of Refs.~\cite{read saleur enlarged symm alg, read saleur assoc alg} on the symmetry algebra of these models. (The operator product expansions of these operators are also constrained by geometrical arguments.) Finally, we show that the sigma model has a novel parity-odd operator whose scaling dimension is fixed by a relation with the loops' winding angles. 

\subsection{Outline}

The theme of the remainder is the collapse transition in various settings, but much of the material is relevant to the $\cp^{N-1}$ model more generally.  Here is an overview:

\noindent
--- Sec.~\ref{background section} reviews the basic models and tools we will need (the first half of this section will be familiar to many readers) with new results presented in subsequent sections.

\noindent
--- Sec.~\ref{relevant operators section} describes the operators in the $\cp^{N-1}$ model that are most important for the discussion of collapse. A full demonstration that these are the only important operators  is deferred until Sec.~\ref{operators section}.

\noindent
--- Sec.~\ref{honeycomb stability section} shows that the archetypal honeycomb model, and by extension any model in the DS universality class, is stable to arbitrary perturbations of the interactions. 

\noindent
--- Sec.~\ref{section on additional perturbation in square lattice model} considers a well-known model  on the square lattice (which we refer to as Model T), in order to resolve an apparent paradox about the stability of the DS point.

\noindent
--- Sec.~\ref{crossings section} argues that models with crossings (Fig.\ref{withwithoutcrossings}, right) will have non-DS exponents and discusses some other features of models with crossings (the special case of `smart walks'). 

\noindent
--- Sec.~\ref{operators section} uses simple geometrical arguments to pin down the scaling dimensions of some interesting operators in the $\cp^{N-1}$ model (or its supersymmetric cousin the $\cp^{N+k-1|k}$ model), specifically  odd-parity variants of the two- and four-leg operators. This also confirms that our classfication of perturbations in the polymer problem is complete.

\noindent
--- Technical results necessary for Secs.~\ref{honeycomb stability section}--\ref{operators section} are given in the appendices, including the lattice mappings that underlie our analysis of perturbations, and  an aspect of the supersymmetric formulation.

\section{Background:\\models and field theories}
\label{background section}

In this section we review the polymer models we consider and their relations with loop gases and field theory.

\subsection{Honeycomb Model}
\label{honeycomb model intro}

 Usually models for a single polymer can be thought of as loop gases in the limit where the loop fugacity, $n$, tends to zero \cite{replica for loops}. (As usual it will be convenient to consider a closed ring polymer rather than an open chain.) The unusual feature of solvable models in the DS universality class is they allow a different type of mapping which is instead between the polymer model and a loop gas at fugacity ${N=1}$. The loops in this gas are essentially cluster boundaries in critical percolation. The correspondence is that the Boltzmann weight of a given polymer conformation is proportional to the probability of a loop with that conformation appearing in the loop gas.
 
 The model of Refs.~\cite{duplantier saleur polymer, coniglio jan et ap} for the collapse transition maps to the much-studied  gas of nonintersecting loops on the honeycomb lattice \cite{O(n) lattice magnet1, O(n) lattice magnet2}:
\be \label{honeycomb Z}
Z_\text{honeycomb} = 
\sum_{\substack{\text{loop}\\ \text{configs}}}  x^\text{length} N^\text{no. loops}
=
\sum_{\substack{\text{coloured}\\ \text{loop configs}}}  x^\text{length}.
\ee
`Length' is the total length of the loops. For the second equality we have assumed $N$ to an integer, allowing us to obtain  the fugacity $N$  by summing over $N$ possible colours for each loop.  We will be interested in the `dense phase' (i.e. $x$ larger than a critical value), in particular at $x=1$ and $N=1$.

\begin{figure}[t]
 \begin{center}
       \includegraphics[width=0.45\linewidth]{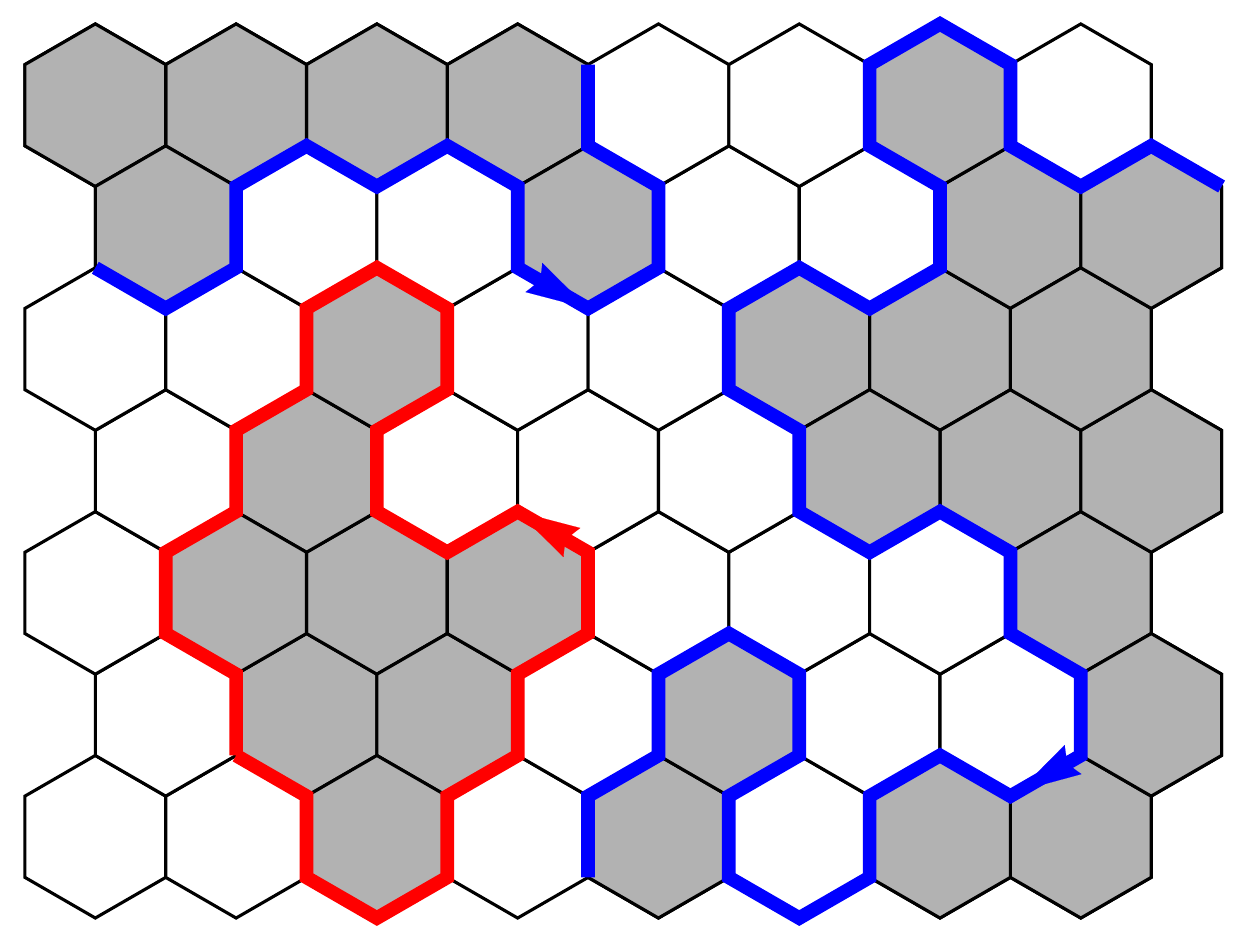}
  \includegraphics[width=0.074\linewidth]{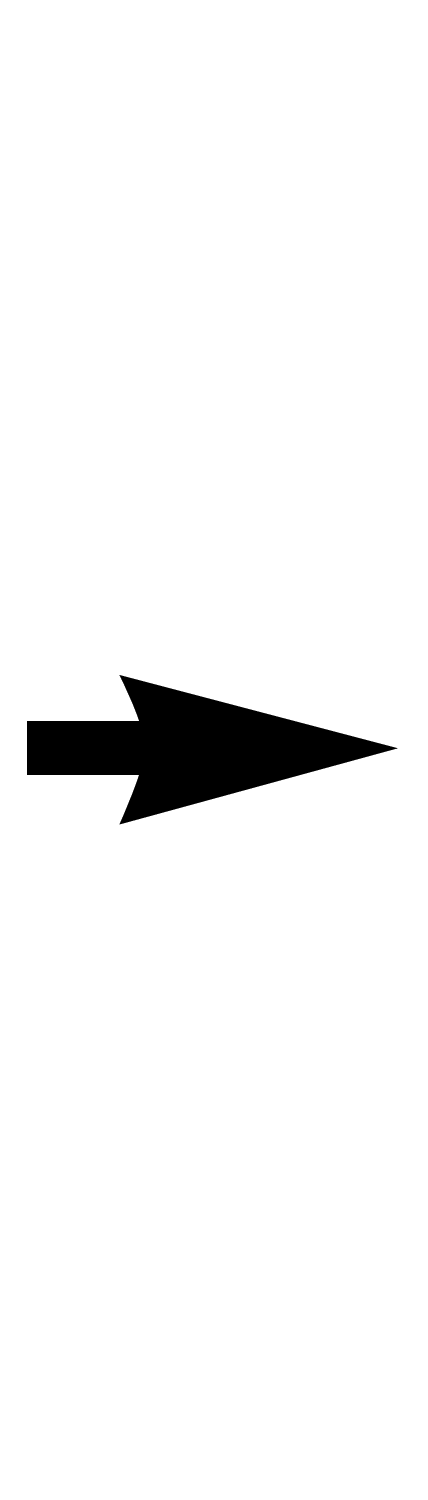}
   \includegraphics[width=0.45\linewidth]{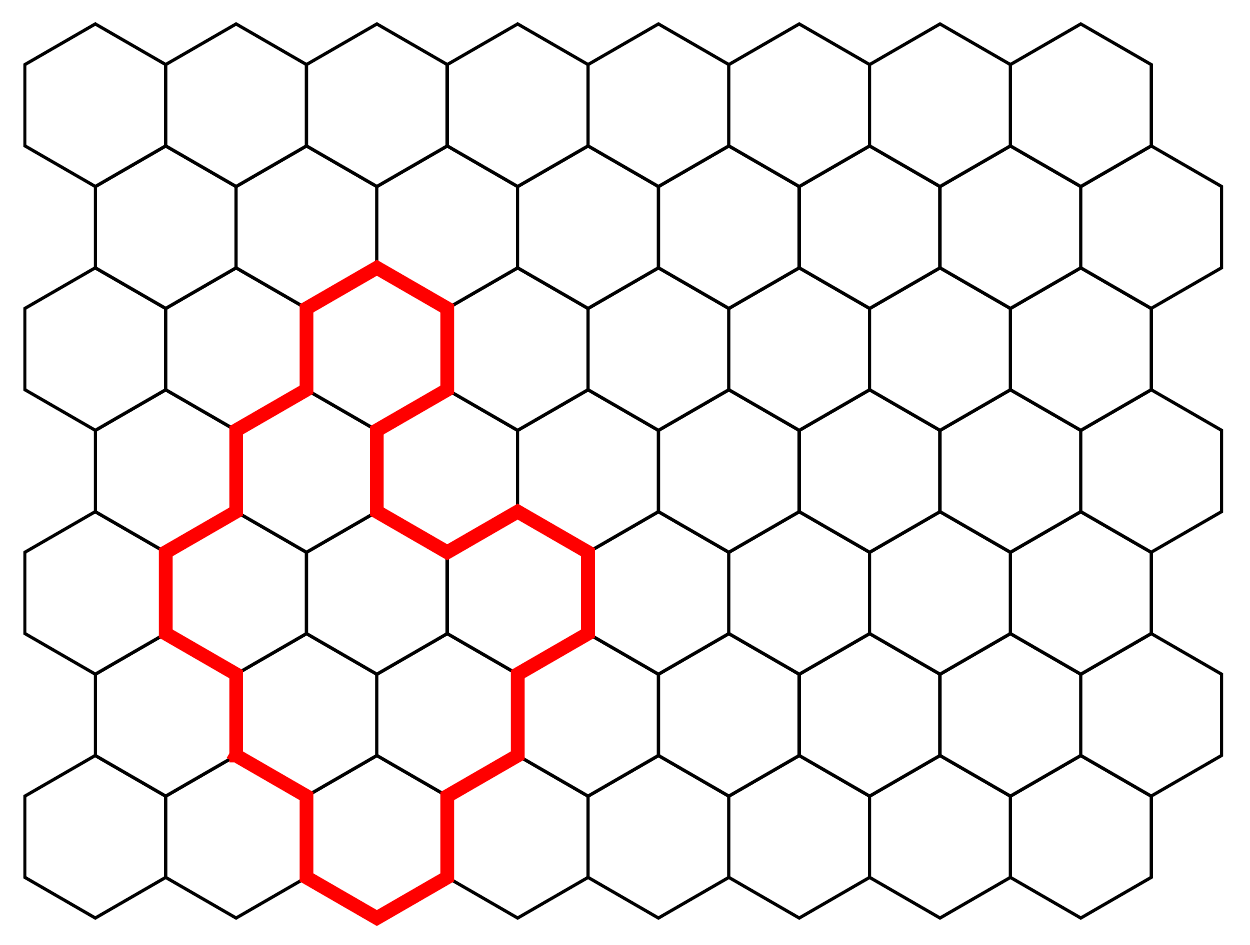}
 \end{center}
 \caption{Correspondence between loop gas on honeycomb lattice (percolation) and polymer model. A randomly chosen loop from the former is statistically equivalent to a ring polymer governed by the Boltzmann weight in Eq.~\ref{honeycomb polymer partition function}.}
 \label{looppolcorrespondencehoneycomb}
\end{figure}

It is useful to regard the loops as cluster boundaries, as in  Fig.~\ref{looppolcorrespondencehoneycomb}, left. Given a loop configuration, the colouring of the hexagons is unique up to a global exchange of white and black: we may for example sum over both choices, which simply multiplies $Z_\text{honeycomb}$  by two. Viewing the loops as cluster boundaries shows that there is a natural convention for \textit{orienting} them: we declare that the loops encircle black clusters in an anticlockwise direction. The fact that the loops in Eq.~\ref{honeycomb Z} are `secretly' oriented has crucial consequences for the continuum theory \cite{read saleur exact spectra}. Viewing the loops as cluster boundaries also shows that at $N=1$ and $x=1$ the above loop gas is nothing but uncorrelated site percolation on the triangular lattice, which is critical  since black and white hexagons are equiprobable.

The above theory at $N=1$ describes a soup of many loops, rather than a single polymer. However at $N=1$, $x=1$ there is a well-known mapping to a partition function for the latter. Crudely, the point is  that a loop picked at random from the loop gas (Fig.~\ref{looppolcorrespondencehoneycomb}) is statistically equivalent to a ring polymer with certain  interactions. (These interactions are local, thanks to the short range correlations in the percolation problem.) Taking a system on a finite lattice, say with periodic boundary conditions, the polymer partition function is:
\be\label{honeycomb polymer partition function}
Z_\text{polymer} =  \sum_{\substack{\text{polymer}\\ \text{configs}}} \lf \f{1}{2} \ri^\text{no. hexagons visited by polymer}.
\ee
The weight may be seen as the combination of a weight per unit length together with attractive interactions of a certain kind (for a given length, more compact configurations are favoured since they visit fewer hexagons). These interactions are such that the polymer is tuned to the collapse point. For example, the mapping to the loop gas implies that the fractal dimension of the polymer is $d_f = 7/4$ \cite{duplantier saleur percolation}, which is in between the self-avoiding walk value ($d_f=4/3$) and the value in the collapsed phase ($d_f=2$).

The introduction of the loop colours in Eq.~\ref{honeycomb Z} gives a useful way of formalising the connection between the gas of many loops and the polymer model \cite{Cardy n+n' trick}. We write 
\be
N = 1+n
\ee
and label the $N$ possible colours for each loop by
\be
a=0,\ldots, n.
\ee
We distinguish loops  of colour $a=0$, which we refer to as  `background' loops, from loops of colour $a=1,\ldots, n$ which we refer to as polymers. Informally, the point is that `integrating out' the background loops in a configuration with a single polymer gives the desired weight in Eq.~\ref{honeycomb polymer partition function}. And since each polymer then has a statistical weight $n$, we can use a replica-like limit $n\rightarrow  0$ to isolate configurations with a single polymer. 

Explicitly, expanding $Z_\text{honeycomb}$  in $n$ gives
\be
Z_\text{honeycomb} = \Bigg( \sum_{\substack{\text{background}\\ \text{loop configs}}} 1  \Bigg)
+  n  \Bigg( \sum_{\substack{\text{configs with}\\ \text{1 polymer}}} 1\Bigg) + \ldots
\ee
The first term is proportional to the sum over percolation configurations. Absorbing this trivial constant into the definition of $Z_\text{honeycomb}$, and performing the sum over the configurations of the background loops in the second term,
\be
Z_\text{honeycomb} = 1 
+  n  \, Z_\text{polymer} + \ldots
\ee

$Z_\text{polymer}$ is the first  $\theta$-point model we will consider. More generally, we also wish to consider the space of models \textit{close} to this one. We will show that the DS $\theta$-point behaviour of this model is robust --- the universality class of the collapse transition remains the same if we slightly change the form of the interactions. If we wish to consider the introduction of crossings this model is not very convenient, so we will be led to consider models on the square lattice.

\subsection{Square Lattice Model (`Model T')}

\begin{figure}[t]
 \begin{center}
    \includegraphics[width=0.45\linewidth]{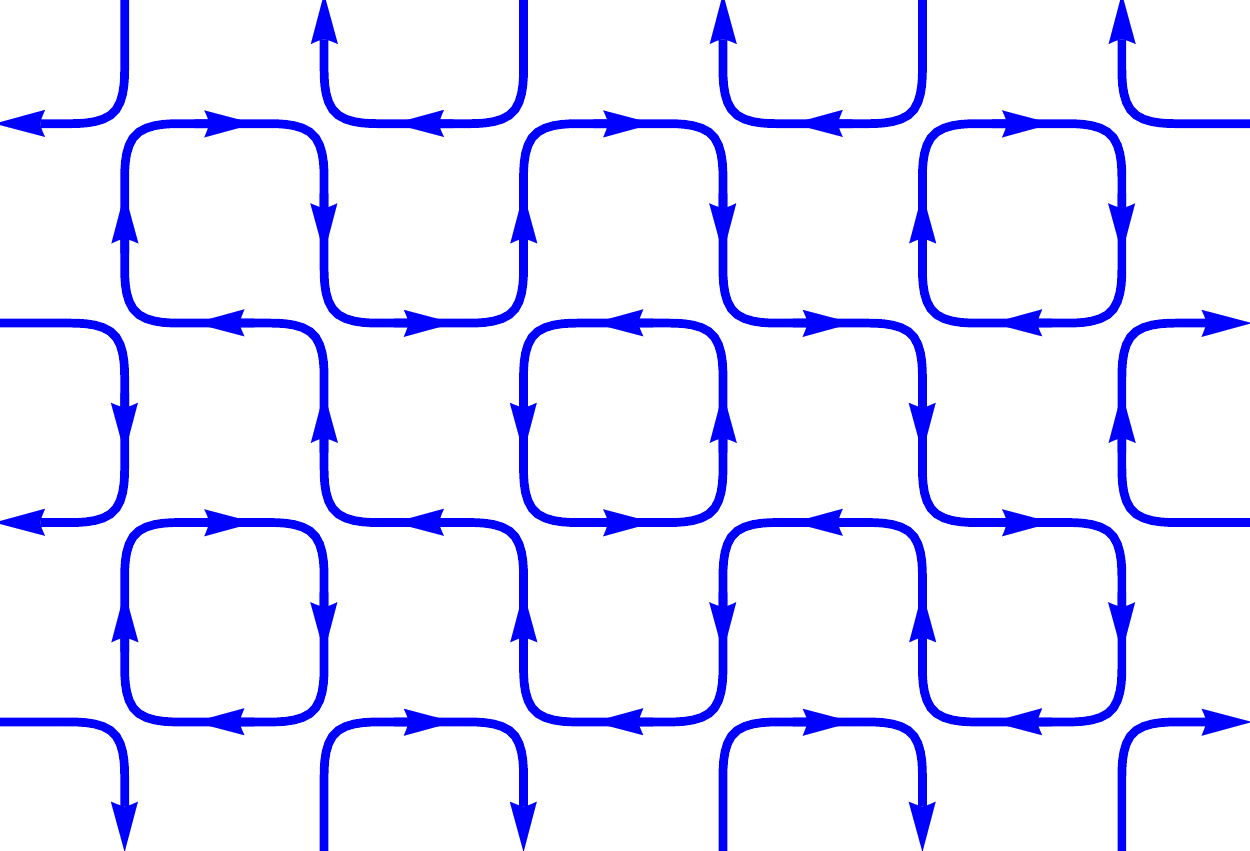}
  \includegraphics[width=0.074\linewidth]{arrowfigure}
   \includegraphics[width=0.45\linewidth]{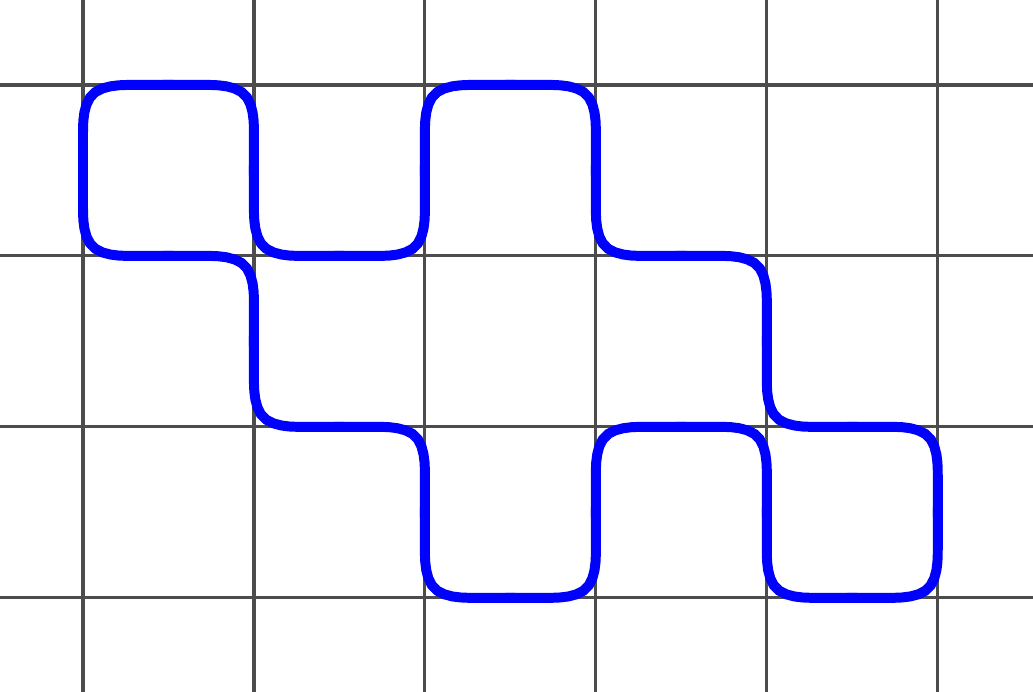}
 \end{center}
 \caption{Correspondence between the completely-packed loop model and polymer model. A randomly chosen loop from the former is statistically equivalent to a ring polymer governed by the Boltzmann weight in Eq.~\ref{CPL polymer partition function}. (The orientations assigned to the links in the left figure are fixed, \textit{not} fluctuating degrees of freedom.)}
 \label{looppolcorrespondence}
\end{figure}

\begin{figure}[b]
\includegraphics[width=0.65\linewidth]{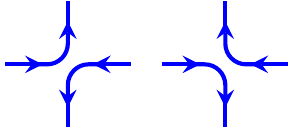}
 \caption{In the completely packed loop model, each node has two possible configurations.}
 \label{nodes}
\end{figure}

The second model derives from the well-known completely-packed loop model on the square lattice (Fig.~\ref{looppolcorrespondence}). Configurations are generated by choosing the pairing of links at each node (Fig.~\ref{nodes}). The partition function is:
\be\label{CPL partition function}
Z_\text{CPL} = \sum_{\substack{\text{loop}\\ \text{configs}}} N^\text{no. loops} 
=\sum_{\substack{\text{coloured}\\ \text{loop configs}}}1.
\ee
Loops in this model always turn at nodes. Therefore if we assign \emph{fixed} directions to the links of the square lattice by the arrow convention in Fig.~\ref{looppolcorrespondence}, the loops acquire consistent orientations.  (This oriented square lattice is known as the L-lattice.) The loop gas is again in the dense phase, and the loops have the same universal properties as those in the honeycomb model. When $N=1$ there is again a correspondence with a polymer model. A loop picked at random from the gas (Fig.~\ref{looppolcorrespondence}, right) is governed by the effective `polymer' partition function:
\be\label{CPL polymer partition function}
Z_\text{model T} =  \sum_{\substack{\text{polymer}\\ \text{configs}}} \lf \f{1}{2} \ri^\text{no. nodes visited}.
\ee
The configurations appearing in the partition function are constrained to {Turn} at each node (Fig.~\ref{looppolcorrespondence}, right) so we refer to this as Model T. Later on we will relax this constraint. This model is well known \cite{blote nienhuis, blote batchelor nienhuis 98, bradley, prellberg owczarek 94}. 

The large-scale properties of a polymer ring in this model are identical to those in Eq.~\ref{honeycomb polymer partition function}. However we will \textit{not} refer to this model as being in the DS universality class. This is because some universal properties differ, despite the fact that the same field theory applies in each case. This will be discussed below (Ref.~\ref{discussion of fate of model T}). In particular correlators for \textit{open} chains are related to field theory correlators in slightly different ways in the two cases, and as a result the entropic exponent $\gamma$ for the partition function of an {open} chain takes a different value in Model T and the honeycomb model \cite{bradley, prellberg owczarek 94}.

\subsection{Sigma Model for Loop Gases}
\label{intro to sigma model}

Loop gases with fugacity $N$ map to nonlinear sigma models for $N$--component fields.  The  best-known example of this is the relationship between the honeycomb model of Eq.~\ref{honeycomb Z} and the high-temperature expansion of a modified  $\mathrm{O}(N)$ lattice magnet \cite{O(n) lattice magnet1, O(n) lattice magnet2}. However the true global symmetry in these models is in fact larger, namely $\SU(N)$, as a result of the fact that the loops do not cross \cite{read saleur exact spectra,  affleck loops}. In the next subsection we will see explicitly how this arises on the lattice. Heuristically, the key point is that in the models without crossings there are natural prescriptions, discussed above, for orienting the loops. The appearance of oriented loops signals that we should be working with a \emph{complex} $N$-component field,
\ba
\zb&=(Z_0, \ldots, Z_{N-1}),&
|\zb|^2=1.
\end{align}
One may think of this as follows. If we treat the 2D space as the Euclidean spacetime for a 1+1D quantum problem, the theory with the complex field describes $N$ `colours' of charged bosons, labelled $a=0,\ldots n$. The loops are simply the worldlines of these bosons. In addition to the colour index labelling the species, they carry an orientation which distinguishes particles from antiparticles.

The appropriate field theory for $\zb$ turns out to have the $\mathrm{U}(1)$ gauge symmetry $\zb(x) \rightarrow e^{i\phi(x)} \zb(x)$ \cite{read saleur exact spectra,  affleck loops, vortex lines}. (This is related to the fact that the orientation of a given loop in  Fig.~\ref{looppolcorrespondence}, left, is not free to fluctuate.) Therefore it is useful to introduce the gauge-invariant field
\be\label{definition of Q}
Q_{ab} = Z_a Z_b^* -N^{-1} \delta_{ab}.
\ee
The traceless matrix $Q$ parametrizes complex projective space, $\cp^{N-1}$ (and satisfies a nonlinear constraint, since $|\zb|^2 = 1$). The field theory describing the nonintersecting loop gas is the $\cp^{N-1}$ nonlinear sigma model with a topological `$\Theta$' term \cite{read saleur exact spectra}:
\be\label{cpn-1 lagrangian}
\LL_{\cp^{N-1}} = \f{K}{2} \tr \, (\nabla Q)^2 + \f{\Theta}{2\pi} \epsilon_{\mu \nu} \tr Q \nabla_\mu Q \nabla_\nu Q.
\ee
The coefficient $\Theta$ is equal to $\pi$ \cite{theta pi footnote}. This sigma model  flows, for sufficiently large bare stiffness $K$ and for $N\leq 2$, to a nontrivial fixed point which describes the dense phase of the loop gas.  The regime of interest to us is $N=1+n$, with $n$ infinitesimal, so $N$ must be treated as variable in the spirit of the replica trick. An alternative is to formulate a supersymmetric version of the sigma model \cite{read saleur exact spectra}: for our purposes the two approaches are equivalent.

The sigma model captures correlation functions in the loop gases, and by extension in the polymer models. It is useful to keep in mind the heuristic picture of the loops as worldlines of $\zb$. So, for example, the operator $Q_{12} = Z_1 Z_2^*$ absorbs an incoming worldline of colour index $a=1$ and emits an outgoing one of colour $a=2$, as illustrated in Fig.~\ref{2and4legfig}. In the next subsection we will make this more precise on the lattice.

Recall the distinction between background loops ($a=0$) and polymer loops ($a=1,\ldots, n$). We make a corresponding splitting of the components of $\zb$,
\ba
\vec Z & = (Z_0, \vec Z_\perp), & 
\vec Z_\perp & = (Z_1, \ldots, Z_n),
\end{align}
with worldlines of $Z_0$ and $\zb_\perp$ corresponding to background and polymer loops respectively. $\vec Z_\perp$  has a vanishing number of components in the limit of interest, namely $N\rightarrow 1$ or $n\rightarrow 0$. `Watermelon' correlation functions for the polymer may be expressed in terms of $\zb_\perp$.

The field theory $\LL_{\cp^{N-1}}$ is appropriate to the polymer model $Z_\text{polymer}$, which derives from a loop gas in which the polymer and background loops are on exactly the same footing. But a general perturbation of the Boltzmann weight for the polymer will --- when translated back to the loop gas --- break the symmetry between the polymer and the background loops \cite{loops with crossings}. Correspondingly, the Lagrangian will be perturbed by operators $\OO_i$ which reduce the $\mathrm{SU}(N)$ symmetry to something smaller:
\be
\LL = \LL_{\cp^{N-1}} + \sum_i \lambda_i \OO_i.
\ee

\subsection{Lattice Field Theories}
\label{lattice partition functions intro}

To make the connection between the loop gases and the sigma model concrete, we will need lattice field theories which (I) map exactly to the loop gas, and (II) turn into the sigma model upon coarse-graining \cite{cpn loops short, loops with crossings}.

\subsubsection{Completely-Packed Model}

First consider the completely-packed model on the square lattice, Eq.~\ref{CPL partition function} (see Refs.~\cite{cpn loops long, deconfined paper, loops with crossings} for more detail). We take a model with $N$-component complex vectors $\zb$ located on the links of the lattice, with fixed length $|\zb|^2 = N$. The Boltzmann weight is a product over terms for each node. Denoting the two outgoing links at a given node by $o$, $o'$ and the two incoming links by $i$, $i'$,
\be \label{lattice cpn-1 model}
Z_\text{CPL} = \Tr \prod_\text{nodes} \lf  (\zb_o^\dag \zb_i) (\zb_{o'}^\dag \zb_{i'}) +  (\zb_{o'}^\dag \zb_i) (\zb_{o}^\dag \zb_{i'})  \ri.
\ee
`$\Tr$' denotes the integral over the $\zb$s with the length constraint. Note that the two terms at each node correspond to the two ways of pairing up the links at that node shown in Fig.~\ref{nodes}. Therefore expanding out the product over nodes generates the sum over loop configurations, with each loop decorated with a product of $\zb^\dag \zb$ factors. In a loose notation where the links on a given loop are denoted $1\ldots,\ell$ as we go around the loop in the direction of its orientation, we have
\be
Z_\text{CPL} = \sum_\text{configs} \, \prod_\text{loops}   \Tr
\lf \zb_1^\dag \zb_\ell \ri
\ldots
\lf \zb_3^\dag \zb_2 \ri
\lf \zb_2^\dag \zb_1 \ri.
\ee
Using $\Tr Z_a Z_b^* = \delta_{ab}$ to integrate out the $\zb$s, we find that each loop has a single colour index $a$ which is conserved along its length. Therefore Eq.~\ref{lattice cpn-1 model} is equal to the partition function of the loop model, $Z = \sum_\text{coloured loop configs}1$.

The above theory has $\mathrm{SU}(N)$ global symmetry and $\mathrm{U}(1)$ gauge symmetry under independent phase rotations on each link, $\zb_\ell \rightarrow e^{i\phi_\ell}\zb_\ell$. One may show that the continuum limit of this lattice field theory is the $\cp^{N-1}$ Lagrangian of Eq.~\ref{cpn-1 lagrangian} with $\Theta=\pi$ \cite{theta term footnote}. This agrees with the field theory derived for the loop model by first taking an anistropic limit  which maps it to a quantum spin chain \cite{read saleur exact spectra}.

Inserting operators on the links modifies the graphical expansion. For example if we insert $Q_{cd}$ on a link, the integral over $\zb$ on that link is modified from $\Tr Z_a Z_b^* = \delta_{ab}$ to $\Tr Z_a Z_b^* Q_{cd} =(n \delta_{ad}\delta_{bc} - \delta_{ab}\delta_{cd})/ (n+1)$. It follows that inserting $Q_{12}$ forces the colour of the incoming part of the strand passing through the link to be $1$, and the colour of the outgoing segment to be $2$. The correlation function $\< Q_{12} (l) Q_{21}(l)\>$ then contains only configurations in which the links $l$ and $l'$ lie on the same loop. That is, $Q_{ab}$ is a lattice `two-leg' operator.

\subsubsection{Honeycomb Model}
\label{honeycomb model review}

The lattice field theory for the honeycomb model given in Ref.~\cite{vortex lines} (see also Ref.~\cite{affleck loops}) is very similar to the lattice magnet of Nienhuis et al. \cite{O(n) lattice magnet1, O(n) lattice magnet2}, but includes a $\mathrm{U}(1)$ gauge field. The role of this gauge field is to fix the relative orientations of the loops in accordance with the cluster boundary convention in Fig.~\ref{looppolcorrespondencehoneycomb}, which leads to adjacent loops being oppositely oriented.

The spins of the lattice magnet are again complex vectors $\zb = (Z_0, \ldots, Z_{N-1})$ with  length $|\zb|^2 = N$, but are now located at the sites $i$ of the honeycomb lattice. The gauge field is an angular degree of freedom $U_{ij} = e^{i a_{ij}}$ which is located  on the links  (with $U_{ij} = U_{ji}^*$). The partition function we need is
\be\label{lattice gauge theory}
Z = \Tr\hspace{-3mm} \prod_\text{hexagons H} \hspace{-1mm} \bigg( 1 + \prod_{\<ij\> \in \text{H}} U_{ij} 
\bigg)
\prod_{\<ij\>} \lf 
1 + x\, U_{ij} \zb^\dag_i \zb_j + \text{c.c.}
\ri.
\ee
In the product $\prod_{\<ij\> \in \text{H}} U_{ij}$, the links are oriented anticlockwise around the hexagon. 

This model allows a graphical expansion similar to the previous, showing its equivalence with $Z_\text{honeycomb}$ in Eq.~\ref{honeycomb Z}. The graphical expansion involves not only loops (which come from the expansion of the product over links in Eq.~\ref{lattice gauge theory}) but also shaded hexagons (which come from the expansion of the product over hexagons). The shaded hexagons make up the black clusters in Fig.~\ref{looppolcorrespondencehoneycomb}, and the loops are glued to the boundaries of these clusters once we integrate over $U_{ij}$.

Eq.~\ref{lattice gauge theory} has the same gauge and global symmetries as Eq.~\ref{lattice cpn-1 model}. The microscopic field content is different because of the presence of the fluctuating gauge field; but this does not change the coarse-grained Lagrangian. (In fact the continuum sigma model, Eq.~\ref{cpn-1 lagrangian}, admits an equivalent formulation with $\zb$ coupled to a dynamical continuum gauge field $A_\mu$. In this formulation the $\Theta$ term is simply proportional to the integral of the flux $\epsilon_{\mu\nu} \nabla_\mu A_\nu$. This term arises naturally from coarse-graining the first term in Eq.~\ref{lattice gauge theory} \cite{vortex lines}.)

\begin{figure}[t]
\includegraphics[width=0.98\linewidth]{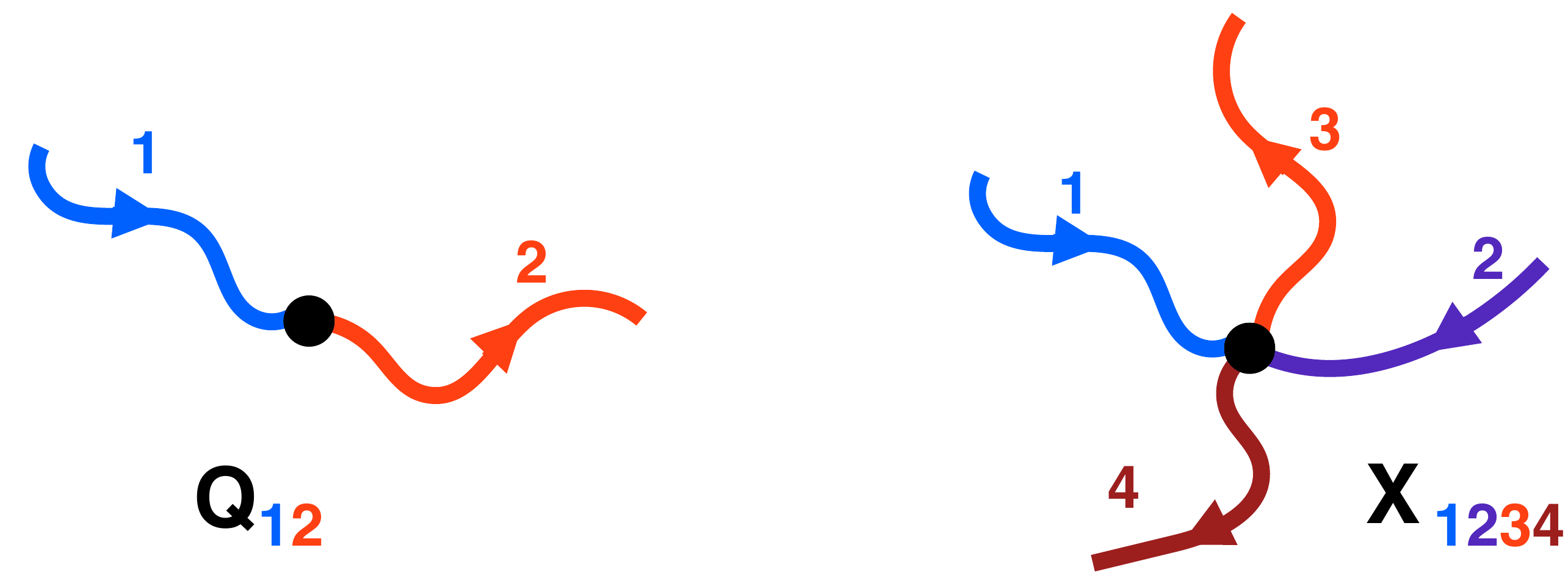}
 \caption{Schematic representation of some components of the two and four-leg operators. These operators enforce vertices with strands of specified colour emanating from them.}
 \label{2and4legfig}
\end{figure}

\section{Relevant Operators}
\label{relevant operators section}

We now return from the lattice to the continuum field theory of Sec.~\ref{intro to sigma model}. The operators pertinent to our discussion of stability will be components of the two- and four-leg operators, both of which are relevant if added to the action. 

As noted above the two-leg operator is essentially the matrix $Q_{ab}$ defined in Eq.~\ref{definition of Q} \cite{operator identification footnote}. It transforms in the adjoint representation of $\mathrm{SU}(N)$, which has dimension $N^2-1$, and its RG eigenvalue in the $N=1$ limit is $y_2=7/4$. A lattice version  of this operator can for example be defined on a link of the completely-packed loop model as discussed above \cite{rotational invariance}, and its two point function is proportional to the probability that two links lie on the same loop.

The four-leg operator comes in two types, as we will discuss in Sec.~\ref{two types of 4 leg operator}, with different behaviour under parity (spatial reflections). Only the parity-even operator is important for the RG flows we consider, since spatial symmetry prevents the parity-odd operator from appearing in the action. This parity-even four-leg operator, $X_{aa'bb'}$, is essentially $Z_a Z_{a'} Z^*_b Z^*_{b'}$, with trace terms subtracted to ensure  it transforms irreducibly under $\mathrm{SU}(N)$ \cite{springer thesis}:
\ba \notag
X_{aa'bb'} =  Z_a Z_{a'} Z^*_b Z^*_{b'} &  -  \f{\delta_{a  b} Z_{a'} Z^*_{b'}  + \text{3 terms} }{N+2}  
\\
&+ \f{\delta_{a b} \delta_{a'b'} + \delta_{a b'} \delta_{a' b}}{(N+1)(N+2)}.
\end{align}
$X_{aa'bb'}$ is symmetric under $a\leftrightarrow a'$ and under $b\leftrightarrow b'$.  Graphically, it is a four-leg vertex with incoming directed lines of colour $a$, $a'$ and outgoing directed lines of colour $b$, $b'$ (Fig.~\ref{2and4legfig}). A lattice version of this operator may be written down in the completely-packed loop model, but will not be needed here. $X$ has RG eigenvalue $y_4 = 3/4$ at $N=1$, and it forms an irreducible representation of $\mathrm{SU}(N)$ whose dimension is 
\be\label{multiplicity of X}
d_X = \f{N^2 (N-1) (N+3)}{4}.
\ee
$Q$ and $X$ above are the \emph{only} operators in the sigma model which are invariant under both spatial rotations and parity and which are relevant at $N=1$. We defer the demonstration of this to Sec.~\ref{operators section}. In order to show that no further relevant or marginal operators  can appear as perturbations to the action, we identify the full set of operators with dimensions $x=x_4$ and $x=2$: we confirm this set is complete using the results of Read and Saleur on the counting of states in the spectrum of the supersymmetric sigma model \cite{read saleur exact spectra}.  We find that no additional perturbations are allowed by parity symmetry.

\begin{figure}[t]
\includegraphics[width=0.98\linewidth]{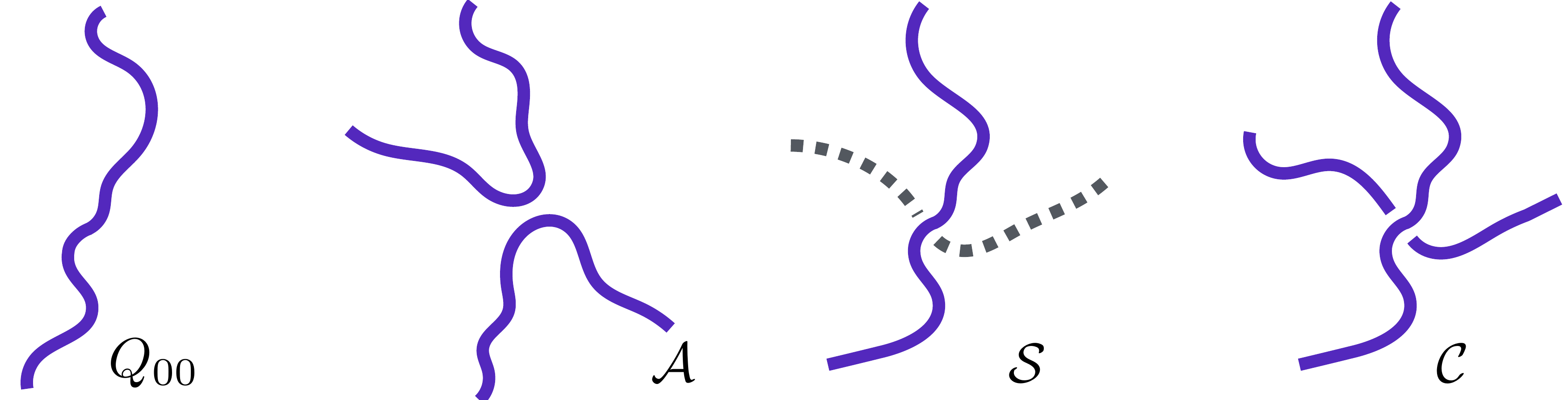}
 \caption{Schematic interpretation of the four relevant perturbations in Eq.~\ref{4oplist}. $Q_{00}$ is the leading perturbation when we change the variable conjugate to the length of the polymer. $\mathcal{A}$ can be independently tuned by modifying the strength of attraction between nearby monomers.  In the loop gas interpretation, $\mathcal{S}$ is a crossing between the polymer strand and a `background' strand (dashed line). In generic models in the DS universality class this perturbation does not play any role. However in Model T, which is fine-tuned in a certain sense, $\mathcal{S}$ corresponds to introducing nodes where the polymer does not turn (straight segments). The perturbation $\mathcal{C}$ arises when the polymer is allowed to cross itself.}
 \label{fourperturbations}
\end{figure}

When we perturb the Boltzmann weight for the polymer the global symmetry will be reduced to a subgroup of $\mathrm{SU}(N)$, and the operators above will split into more than one representation of the reduced symmetry. Four operators will play a role in the discussion of RG flows below:
\ba\label{4oplist}\notag
&Q_{00} = - \sum_{a=1}^n Q_{aa}; &
&\mathcal{A} \equiv - X_{0000} =  -  \sum_{a,b=1}^n X_{abab}; \\
& \mathcal{S} \equiv - \sum_{a=1}^n X_{aa00} + \text{c.c.}; &
&\mathcal{C} \equiv - \sum_{a,b=1}^n X_{aabb}.
\end{align}
The effects of these perturbations are summarised heuristically in Fig.~\ref{fourperturbations}, and will be explained in the following sections.

\section{Perturbing the honeycomb model}
\label{honeycomb stability section}

First we show that the collapse transition in the honeycomb polymer model (Eq.~\ref{honeycomb polymer partition function}) remains in the DS universality class even when the interactions between monomers are slightly perturbed. 

Recall that the polymer is a worldline of the field $\zb_\perp$. Therefore  we might expect that the effect of changing the interactions between monomers will simply be to add local interactions for $\zb_\perp$.  Neglecting terms with derivatives, the perturbed Lagrangian will then be of the form
\be
\LL = \LL_{\cp^{N-1}} + V\big(|\zb_\perp|^2\big),
\ee
where $V$ is an arbitrary potential. This expectation is correct (though see next section). In Appendix~\ref{honeycomb lattice perturbations} we confirm this explicitly  for an arbitrary perturbation of the interactions, using the lattice field theory in Eq.~\ref{lattice gauge theory}.  The resulting symmetry breaking is (recall $N = 1+n$):
\be\label{sun to un}
\mathrm{SU}(N)\, \rightarrow \, \mathrm{U}(n).
\ee
The remaining $\mathrm{U}(n)$ rotates the components of $\zb_\perp$.

As discussed in the previous section, any relevant perturbations allowed by spatial symmetry must be sums of components of $Q$ or $X$, with RG eigenvalue $y_2=7/4$ or $y_4=3/4$ respectively.  Taking into account $\mathrm{U}(n)$ symmetry, there are only two linearly independent possibilities:
\be
Q_{00}=-  |\zb_\perp|^2 + {n}/{N}
\qquad 
\text{and}
\qquad
\attr \equiv   -  {\sum}_{a,b=1}^n X_{abab}.
\ee
These appear when for example we change the monomer fugacity (weight per unit length) or the strength of self-attraction (`$\mathcal{A}$' stands for `attraction'). Increasing the monomer fugacity  favours links of colour $a=1,\ldots, n$ over those of colour $a=0$, so naturally generates a positive `mass' for $Z_0$, or equivalently a negative mass for $\zb_\perp$. In a coarse-grained picture, increasing the polymer's self-attraction means increasing the weight for a meeting of four polymer legs, explaining the appearance of the four-leg operator $X_{abab}$ with colour indices $a$ and $b$ both greater than zero. Note that, by virtue of the tracelessness of $Q$ and $X$, operators like $\sum_{a>0} Q_{aa}$, or $\sum_{a>0} X_{a0a0}$, or $X_{0000}$ are linearly related to those above, so do not constitute independent RG directions.

Two RG--relevant directions is the right number for the $\Theta$ point. One relevant perturbation is automatically tuned to zero by taking the polymer to be long \cite{cardy book}, and the other must be varied to reach the collapse point. Therefore  the above shows that the DS behaviour is robust for nonintersecting models on the honeycomb lattice. (The generic form of such a model is given in App.~\ref{honeycomb lattice perturbations}.) This conclusion is consistent with (and explains) early numerical transfer matrix calculations \cite{duplantier saleur numerics}, which investigated several perturbations of the honeycomb Boltzmann weights and found that that the exponents remained the same to within numerical accuracy.

In order to infer that the DS universality class is robust to \textit{any} local perturbations (with the exception of allowing the polymer to cross itself), we should  check that there is no fine--tuning hidden in the choice of lattice. Fortunately, the lattice gauge theory representation makes clear that  $\mathrm{U}(n)$ is retained so long as we do not allow the polymer to cross itself, regardless of the choice of lattice \cite{bead spring footnote}; and so long as $\mathrm{U}(n)$ symmetry is retained, only the two relevant perturbations discussed above can appear in the action. Therefore the DS universal behaviour is generic so long as crossings are forbidden. Physically, only one parameter needs to be varied to reach a collapse transition in this universality class.

This resolves a longstanding theoretical question about this transition. One of the reasons why the stability of the DS fixed point has previously  been a vexed question is that the traditional Coulomb gas \cite{cg reviews} approach to loop models hides the $\mathrm{SU}(N)$ symmetry: this prevents us from being able to classify and count perturbations. (Other sources of confusion have included the assumption that the DS exponents are the same as those of de Gennes' tricritical $\mathrm{O}(n\rightarrow 0)$ model, which we will argue is not the case,  and the existence of various other solvable fixed points that were candidates for the $\theta$ point \cite{blote nienhuis, warnaar, saleur susy, cardy susy}.)  The sigma model is  the right formulation, as we have seen.  However, even in this formulation it is easy to be misled as we will see in a moment.  

The potentially confusing point is related to the fact that the perturbation $\mathcal{S}$ does not appear in the above analysis. This corresponds to a crossing between a polymer and a background strand, and effects a more drastic symmetry breaking than that in Eq.~\ref{sun to un} (Sec.~\ref{section on additional perturbation in square lattice model}).  If this  perturbation was allowed, it would destabilize the DS universal behaviour. This perturbation does \textit{not} appear when we perturb models that are truly in the DS universality class like that above, as we have seen. But we will now see that it does appear when  we perturb Model T  (the model with turns at every node).  This additional perturbation means that Model T is fine-tuned, as first argued by Blote and Nienhuis \cite{blote nienhuis, blote batchelor nienhuis 98}. It also occupies a different position in the phase diagram to the `true ' DS fixed point, and strictly it should be regarded as a distinct universality class. The apparent paradox --- that the two models have different numbers of relevant perturbations despite being described by the same  $\cp^{N-1}$ field theory --- will be resolved in the next section, where we discuss models without crossings on the square lattice. Then in Sec.~\ref{crossings section} we use the square lattice to introduce crossings, which is awkward to do on the honeycomb.

\section{The square lattice model: paradox \& resolution}
\label{section on additional perturbation in square lattice model}

In the square lattice model of Eq.~\ref{CPL polymer partition function} (Model T), the polymer is constrained to turn at each node. The `gentlest' perturbations of this model change the interactions while  retaining this constraint. For this class of models, the story is the same as the previous section: the only relevant operators that arise are $Q_{00}$ and $\attr$ (Appendix~\ref{app perturbations of CPL}) and the universal behaviour remains unchanged. However, an additional relevant perturbation arises if we relax the constraint of turning at every node \cite{springer thesis}.

Recall that this polymer model is related to a \textit{completely packed} loop gas, Fig.~\ref{looppolcorrespondence}. In the language of the loop gas, a non-turning node is a crossing between a polymer strand and a background strand.  This is a vertex  resembling Fig.~\ref{2and4legfig} (Right) where the two outgoing links have colour index $a=0$ (background) and the two incoming links have $a>0$ (polymer), or vice versa. The new perturbation is denoted $\mathcal{S}$,
\ba\label{Ostraight}
\stra & = - \lf \mathcal{N} + \mathcal{N}^* \ri,
& 
\mathcal{N} & \equiv \sum_{a=1}^n X_{aa00} = \lf Z_0^*\ri^{2} \,  \vec Z_\perp^{\,T} \vec Z_\perp .
\end{align}
The appearance of this perturbation can be confirmed directly using the lattice field theory representation of the loop gas (Appendix.~\ref{app perturbations of CPL}).  The operator $\mathcal{S}$ effects the symmetry breaking
\be\label{straight segment symm breaking}
\mathrm{SU}(N)\rightarrow \mathrm{O}(n),
\ee
where the remaining symmetry rotates $\vec Z_\perp$. 

$\stra$ and $\attr$ both derive from the four-leg tensor, but they are linearly independent operators. Therefore, for Model T, the number of relevant directions is \textit{three} when straight segments are allowed. This implies that Model T describes a fine-tuned collapse point. That this model is fine-tuned was originally suggested by Blote and Nienhuis  \cite{blote nienhuis}: the above provides a precise field-theoretic  version of their argument (from the decay of the appropriate correlator) that non-turning nodes should be a relevant perturbation. (The field theory formulation makes it clear that this perturbation is linearly independent of $\mathcal{A}$ at the fixed point.) However it is not correct to infer from the instability of Model T that all the models described by $\cp^{N-1}$ are fine-tuned, as we will discuss. We emphasize that strictly speaking Model T is not in the DS universality class  (Sec.~\ref{discussion of fate of model T}).

The presence of an relevant additional perturbation in this model prompts various questions. First, we have already argued (Sec.~\ref{honeycomb stability section}) that all models in which the polymer does not cross itself have a $\mathrm{U}(n)$ symmetry. At first sight this is in conflict with  Eq.~\ref{straight segment symm breaking}, which says that the $\mathrm{SU}(N)$ symmetry of Model T is broken all the way to $\mathrm{O}(n)$ when the model is perturbed. The resolution is that the perturbed Model T \textit{does} have a $\mathrm{U}(n)$ symmetry, which is revealed by mapping it to a lattice field theory in a different way. However, this $\mathrm{U}(n)$ is not a subgroup of the $\mathrm{SU}(N)$ of the unperturbed Model T! (This subtlety arises because the two ways of mapping the model to field theory are not related by a local change of variable.) In order to see this (Sec.~\ref{hidden symmetry section}) it will be convenient  first to introduce a `less peculiar' square lattice model (Sec.~\ref{second square lattice model section}). The latter also gives an explicit example of a model which has non-turning nodes and is in the (true) DS universality class.

Second, the fact that the perturbation  $\mathcal{S}$ appears here makes it surprising at first sight that it does not appear for models in the DS universality class.  The result of Sec.~\ref{honeycomb stability section} is enough to show that it does not appear when we perturb the DS fixed point, but we can nevertheless ask what it would mean to add it to the Lagrangian in that case. We discuss this briefly in Sec.~\ref{nonlocality of S perturbation}: we find that for the true DS models, $\mathcal{S}$ does not correspond to a \textit{local} perturbation of the polymer Boltzmann weight.

Thirdly, it is natural to ask for a heuristic understanding of why Model T differs from models that are truly in the DS universality class. Here the key player is the Ising variable of Refs.~\cite{blote nienhuis, blote batchelor nienhuis 98}. Model T occupies a different position in the phase diagram to the true DS fixed point: it represents a transition into a collapsed phase with an additional lattice dependent `Ising' order. Relatedly, it is natural to ask how models with the same field theory description can be in different universality classes. We discuss this in Sec.~\ref{discussion of fate of model T}. We also briefly discuss the possibilities for what fixed point Model T flows to when it is perturbed with non-turning nodes.

\subsection{A Less Peculiar Square Lattice Model}
\label{second square lattice model section}

This section introduces a square lattice model which is in the (true) DS universality class. This model does not have Model T's peculiar feature of turning at every node. It lends itself to a different field theory mapping which sheds light on the above issues.

The mappings between polymer models and field theories in Sec.~\ref{lattice partition functions intro} started by relating the former to a gas of oriented loops. We have seen two types of convention for doing this. For the honeycomb model, the loops were oriented by viewing them as cluster boundaries, while  for the completely packed model the loops were oriented by assigning fixed directions to the links. The fact that we could  consistently orient the loops in this way relied on fine-tuning in Model T (the absence of non-turning nodes).

We may also consider loop gases on the square lattice that are \textit{not} completely-packed, and associated polymer models. In fact, since the  background loops are not physical degrees of freedom, we may be able  to map a given polymer model to a loop gas (and then to a lattice field theory) in more than one way, and one mapping may reveal a symmetry which is hidden by the other.

\begin{figure}[t]
 \begin{center}
       \includegraphics[width=0.45\linewidth]{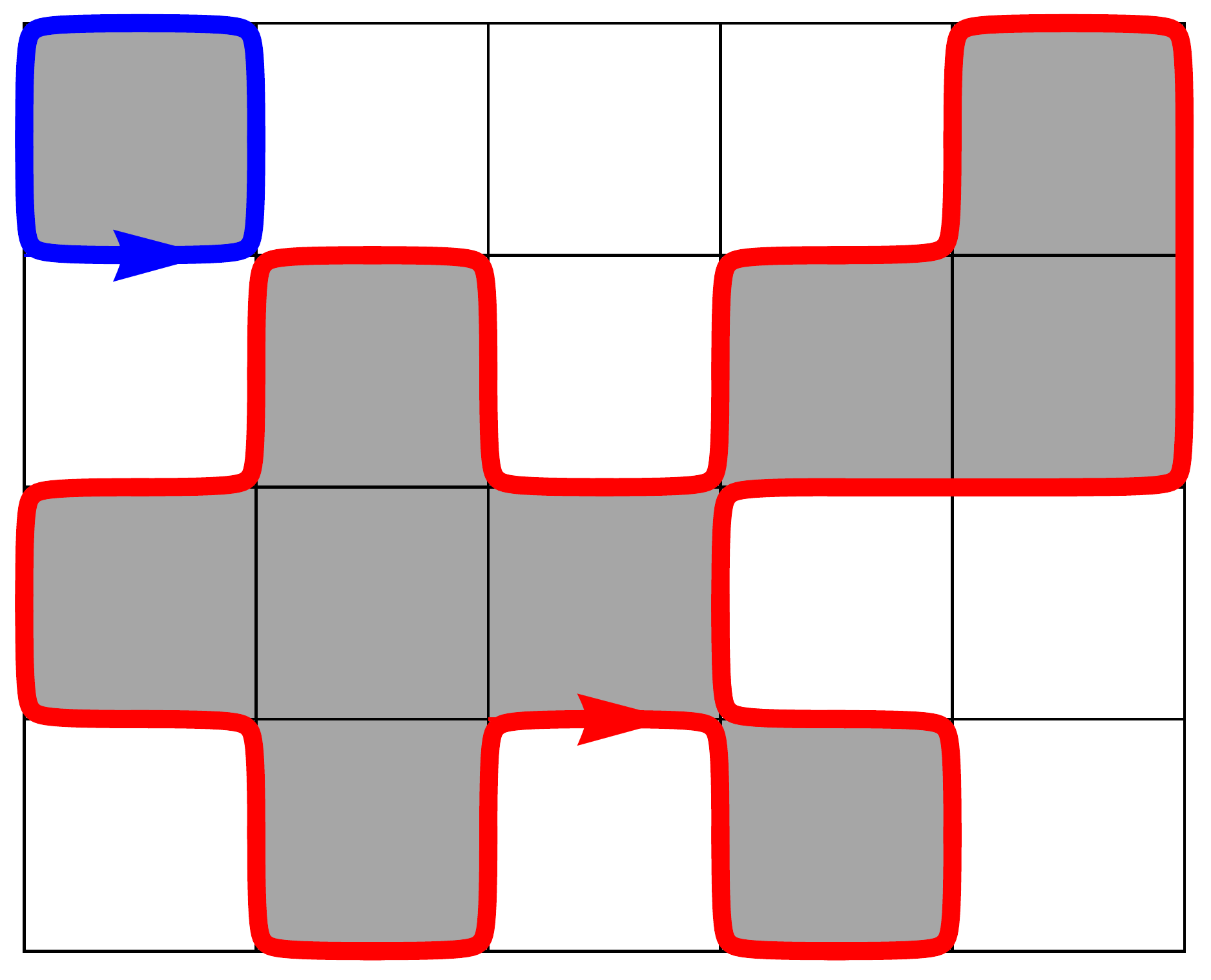}
  \includegraphics[width=0.076\linewidth]{arrowfigure}
   \includegraphics[width=0.45\linewidth]{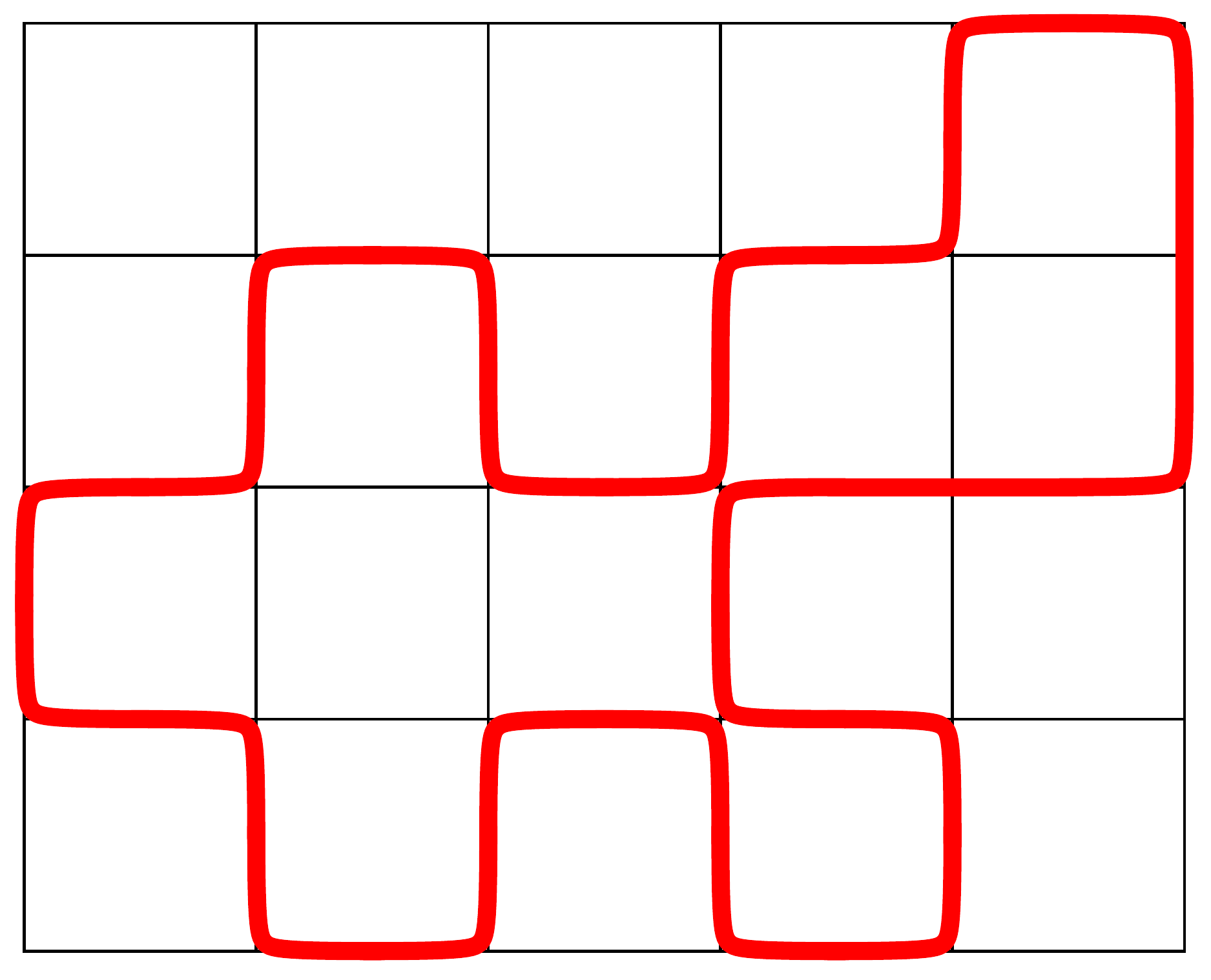}
 \end{center}
 \caption{A second square lattice loop gas (left) that maps to a polymer model (right). In this model the polymer can visit a site twice, but cannot cross. Non-turning nodes are allowed.}
 \label{second square model fig}
\end{figure}

For a specific polymer model  (with non-turning nodes) which is demonstrably in the DS universality class, let us consider  the natural square-lattice analogue of the honeycomb model.  Again we begin with an $N=1$ loop model in which the loops can be viewed as cluster boundaries: see Fig.~\ref{second square model fig}. The only difference with the honeycomb case is that now two clusters can meet at a corner. In this case there are two possible ways to connect up the cluster boundaries (similar to Fig.~\ref{nodes}), which means that a given  configuration of shaded faces can correspond to more than one loop configuration.  The loop gas partition function  is
\be\label{second square lattice model}
Z = \sum_{\substack{\text{loop}\\ \text{configs}}} \alpha^{-T} N^\text{no. loops}.
\ee
Nodes may be visited twice but the loops do not cross. $T$ denotes the number of twice-visited nodes, and $\alpha$ is a constant which we take to be $\alpha=1/2$ when when $N=1$. The loop gas  then maps to a percolation problem in which we (I) colour the faces black or white with equal probability and (II) make random binary choices for how to connect up the cluster boundaries at each  twice-visited node. The fact that the weight of a percolation configuration is shared equally between the two ways of connecting up the clusters at twice-visited nodes gives $\alpha=1/2$. (The non-standard definition of clusters here means that this is different to conventional site percolation on the square lattice. Symmetry between black and white ensures that the present model is critical.)

We can relate this loop gas to a polymer model in the usual way (Sec.~\ref{honeycomb model intro}).  The precise polymer interactions, given in Appendix~\ref{appendix square perc model}, are cumbersome to write down but perfectly local.  The relation with percolation ensures that the polymer is right at its collapse point, and in the DS universality class.

We may also map this model to a lattice gauge theory in an identical manner to the honeycomb model. The continuum limit is again the $\cp^{N-1}$ model at $\Theta=\pi$. As for the honeycomb model, this lattice gauge theory representation can be generalized to allow an \textit{arbitrary} local perturbation to the Boltzmann weight. This is explained in  Appendix~\ref{appendix square perc model}. A convenient intermediate step is to first map the problem to a loop model on a modified lattice, in which each node is replaced by a cluster of trivalent nodes: this ensures that the conformation is specified uniquely by which links are visited, making it easy to write down the interactions in the lattice field theory language.

 The conclusions about stability confirm what we already know from Sec.~\ref{honeycomb stability section}. So long as the polymer cannot cross itself, $\mathrm{U}(n)$ symmetry is retained, and  DS universal behaviour  is robust against (sufficiently weak) perturbations.

\subsection{Hidden $\mathrm{U}(n)$ Symmetry in Non-Crossing Models}
\label{hidden symmetry section}

By a suitable (non-infinitesimal) deformation of the lattice gauge theory representation introduced for the model above, we can in principle describe any non-crossing polymer model on the square lattice while retaining $\mathrm{U}(n)$ symmetry (Appendix~\ref{appendix square perc model}). This includes Model T perturbed by straight segments. 

How do we reconcile this with the $\mathrm{SU}(N)\rightarrow \mathrm{O}(n)$ symmetry breaking that we found in Sec.~\ref{section on additional perturbation in square lattice model}? Both results are correct: the symmetry depends on the way in which we map the polymer model to  field theory, or equivalently on the way in which we introduce the background loops. For Model T, the advantage of the original representation (based on the completely packed loop model) is that it makes the $\mathrm{SU}(N)$ symmetry of the unperturbed Model T manifest. The advantage of the alternative representation is that it makes the $\mathrm{U}(n)$ symmetry of the perturbed model manifest. However this $\mathrm{U}(n)$ should not be regarded as a subgroup of the $\mathrm{SU}(N)$ of the unperturbed Model T, since the two representations involve distinct sets of fields (not  related by  any local transformation). For this reason the alternative representation does not make the $\mathrm{SU}(N)$ symmetry of Model T manifest. On the other hand it does reveal another $\mathrm{SU}(N)$ symmetry at a \textit{different} point in parameter space, namely for the  `less peculiar' model of the previous section. The common feature of the  points at which an $\mathrm{SU}(N)$ symmetry exists is that they map to $N=1$ loop gases. (But differences between these loop gases lead to differences in the polymer models which we touch on in Secs.~\ref{nonlocality of S perturbation},~\ref{discussion of fate of model T}.)

Retaining $\mathrm{U}(n)$ symmetry is enough to ensure that models (without crossings) which are {sufficiently close} to the DS fixed point will flow to it. This includes for example any model which is sufficiently close to the  `less peculiar' model (in which $\mathrm{U}(n)$ is enlarged to $\mathrm{SU}(N)$ microscopically). This does not of course imply that \textit{all} models with $\mathrm{U}(n)$ symmetry lie in the basin of attraction of the DS fixed point.  Therefore we cannot assume that when we perturb Model T with non-turning nodes it will flow to the true DS fixed point. It may flow to a different  fixed point which is also `stable'. We discuss this briefly in Sec.~\ref{discussion of fate of model T}.

\subsection{Absence of Perturbation $\mathcal{S}$ at the (True) DS Fixed Point}
\label{nonlocality of S perturbation}

As a concrete  instance of the DS universality class let us take the `less peculiar'  model on the square lattice (Sec.~\ref{second square lattice model section}). When mapped to field theory appropriately, this is seen to have $\mathrm{SU}(N)$ symmetry. This is broken down to $\mathrm{U}(n)$ when the model is perturbed, and we have seen explicitly that the perturbation $\mathcal{S}$ does not arise. But what happens if we insist on adding this operator to the Lagrangian? 

$\mathcal{S}$  corresponds to a crossing between a polymer strand and a background strand. In the \textit{loop gas} this is a perfectly local perturbation. However, it corresponds to a {non-local} perturbation of the \textit{polymer} model.  To see this, consider (for simplicity) a polymer loop in the shape of a large square with sides of length $L$. Let the weight of this configuration in the polymer partition function be $\mathcal{W}(\lambda_\mathcal{S})$, where $\lambda_\mathcal{S}$ is the weight associated with a crossing between polymer and background strands. We may easily check (using the relation between the loops and percolation) that at small $\lambda_\mathcal{S}$ and large $L$,
\be\label{disallowed taylor expansion}
\mathcal{W}(\lambda_\mathcal{S}) = \mathcal{W}(0) \lf 1 + O(\lambda_\mathcal{S}^2 L^2) \ri.
\ee
The leading correction is $O(\lambda_\mathcal{S}^2L^2)$ for a simple reason: if a background strand enters the polymer loop, it must also leave (giving two $\lambda_\mathcal{S}$ insertions) and there are $O(L)$ choices for both the entry point and the exit point. But we may easily check that a Taylor expansion of this form cannot arise if $\mathcal{W}(\lambda_\mathcal{S})$ is a local Boltzmann weight for the polymer, i.e. a function of the schematic form $\mathcal{W}(\lambda_\mathcal{S})= \exp  \sum_{\vec r}  C_{ r}(\lambda_\mathcal{S})$, where $C_{\vec r}$ is a local term in the Hamiltonian which depends on some finite region around position $\vec r$. Expanding this in $\lambda_\mathcal{S}$ gives 
\ba
\mathcal{W}(\lambda_\mathcal{S})=\mathcal{W}&(0) \Bigg\{
1+ \lambda_\mathcal{S} \sum_{\vec r} C_{\vec r}'(0) \\\notag
&
 + \f{\lambda_\mathcal{S}^2}{2}
\bigg(
\bigg( \sum_{\vec r} C_{\vec r}'(0)\bigg)^2 + \sum_{\vec r} C_{\vec r}''(0)
\bigg)+ \ldots
\Bigg\}.
\end{align}
Generically the leading correction is $O(\lambda_\mathcal{S} L)$. We see that it vanishes only if the $O(\lambda_\mathcal{S}^2 L^2)$ term also vanishes, so an expansion of the form (\ref{disallowed taylor expansion}) is not possible for a local Hamiltonian.

\subsection{Why is Model T Different?}
\label{discussion of fate of model T}

Recall that for models on the square lattice, we may define an Ising variable associated with the polymer \cite{blote batchelor nienhuis 98}. The following definition is equivalent to that of Ref.~\cite{blote batchelor nienhuis 98}. We consider a single polymer loop, which we take to be  consistently oriented along its length. On each link we can then compare the polymer's orientation  with the fixed  link orientation defined by the L lattice (Fig.~\ref{looppolcorrespondence}). We define the Ising-like variable $\sigma_\ell$ on link $\ell$ to be $+1$ if the two orientations agree and $-1$ if they disagree. As we go along the polymer, the domain walls in  $\sigma$  are precisely the non-turning nodes.  

The role of $\sigma$ is simplest  in the phase in which the polymer is dense (but not necessarily completely dense), accessed by increasing its length fugacity beyond the critical value. Since the polymer visits a finite fraction of the sites of the lattice we can define a coarse-grained Ising spin $\sigma(r)$, and the dense phase can be subdivided into two types, depending on whether $\sigma$ is ordered or disordered  \cite{blote nienhuis, blote batchelor nienhuis 98}.  The same is true of the collapsed phase (the collapsed polymer forms a bubble of the dense phase, surrounded by the vacuum).

For Model T, $\sigma_\ell$ is perfectly ordered along the length of the polymer, while for models in the `true' DS universality class, $\sigma_\ell$ is disordered. Heuristically, this `order' in $\sigma_\ell$ is the reason that Model T has an additional RG relevant direction, which corresponds to allowing $\sigma$ to fluctuate. The order in $\sigma$ also implies that Model T lives in a different part of the phase diagram to the generic $\theta$ point  \cite{blote nienhuis, blote batchelor nienhuis 98}. Model T describes the transition between the extended phase and the Ising-ordered collapsed phase with $\<\sigma\>\neq 0$ (which is what we access by perturbing Model T with an an additional attraction  \cite{ising domain wall footnote}). For the models in the DS universality class, however, infinitesimal perturbations will  instead lead to the Ising-disordered collapsed phase with $\<\sigma\>=0$. The very existence of the Ising-\textit{ordered} phase is of course a lattice artifact \cite{blote nienhuis, blote batchelor nienhuis 98}. 

In what sense are the universal properties of Model T  different from those of the true DS point? It shares the same field theory description and many of the same exponents (the watermelon exponents for an even number of legs are the same). The correlations of the Ising order parameter are one difference. More importantly, the exponent $\gamma$ governing the scaling of the partition function for an {open} chain is different for the two fixed points \cite{bradley}. Both fixed points are described by the $\cp^{N-1}$ model, but in order to fully specify the universality class we need some additional information about how  correlators in the $\cp^{N-1}$ model map to correlators for the polymer. This differs slightly for Model T since the mapping arises from a completely packed loop gas. We have seen an example of this in Sec.~\ref{nonlocality of S perturbation}, where an operator in $\cp^{N-1}$ mapped to a local object for the polymer in one case but not the other. The interpretation of the polymer one-leg operator  in terms of $\cp^{n-1}$ is also different in the two cases, reflecting the well known fact that in Model T a one-leg operator for the polymer corresponds to a 2-leg operator in the loop gas.

As an aside, let us consider a simpler example of the fact that the same field theory can be compatible with two slightly different universality classes. These are the \textit{dense} polymer phases with and without Ising order. Here `order' for $\sigma$ has a more straightforward meaning than at the collapse point, since the polymer visits a finite fraction of the links on the lattice. This case is also simpler because we can stick with a single mapping from the polymer to field theory instead of worrying about two.

Consider the `less peculiar' polymer model of Sec.~\ref{second square lattice model section} and its mapping to the $\cp^{N-1}$ model via the \textit{incompletely}-packed loop gas and lattice gauge theory. We  increase the polymer's length fugacity (i.e. perturb with  $Q_{00}$) so that  we enter a dense polymer phase. The field $Z_0$ becomes massive, and we can integrate it out \cite{integrate out footnote}. This leaves us with the $\cp^{n-1}$ sigma model, $n\rightarrow0$, which is the expected  description of a  dense polymer  \cite{read saleur exact spectra, candu et al}.  Initially we are in the Ising-\textit{disordered} dense phase, $\<\sigma\>=0$. 

By decreasing the weight of nonturning nodes, we may drive the transition into the Ising-ordered dense phase. In both  phases, the fluctuations of $\sigma$ are massive, and decoupled from the $\cp^{n-1}$ sector. (The two sectors are decoupled even at the Ising transition \cite{blote nienhuis}.) We might think that the scaling of the watermelon correlators will be the same in the two phases, since the nontrivial $\cp^{n-1}$ sector has not undergone a phase transition  \cite{cts exponents footnote}, but this is not quite true. Consider  the one-leg operator for the polymer. This acts both in the $\cp^{n-1}$ sector and in the Ising sector. In the Ising sector, the endpoint of an open chain should be viewed as a twist or `disorder' operator --- i.e the endpoint of a branch cut --- for $\sigma$. This convention is necessary in order to ensure that the  interactions between the $\sigma$ values of different parts of the chain are effectively local: for example two parts of the chain can only visit the same node if they have the same value of $\sigma$. (In the $\cp^{n-1}$ sector, we cannot write the one-leg operator simply as $Z_a$, since that is not gauge invariant, but one can argue from the lattice gauge theory that the one-leg operator can be incorporated as a twist defect \cite{twist footnote}.) When $\sigma$ is disordered, the branch cut in $\sigma$ costs only $O(1)$ free energy, so the scaling of the one-leg correlator is determined solely by the $\cp^{n-1}$ sector, giving a power law decay. However when $\sigma$ is ordered the branch cut costs a free energy proportional to its length. Therefore we expect that in this phase the one-leg correlator scales exponentially with length and the two endpoints of an open chain are confined together. 

Returning to the collapse transition in the regime where $\sigma$ is playing a role, the nature of the RG flows between the various fixed points is not yet clear. (See Ref.~\cite{vernier new look} for a related discussion.) In particular, what universality class of collapse transition do we get when we slightly perturb Model T with non-turning nodes? A priori there are two possible scenarios:

(I) We could flow from Model T to the true DS universality class. This would be rather unusual, because it would be a flow from one fixed point described by $\cp^0$ to another fixed point also described by $\cp^0$, with the interpretation of the background loops changing during the flow. In this scenario, the perturbation would destroy the Ising `order' along the length of the polymer, but would leave the statistics of a large ring unchanged. The statistics of an open chain would change, since the exponent $\gamma$ is different in the two cases. This scenario would leave the role of the ``branch 3'' fixed point mentioned below somewhat mysterious, however.

(II) We could  flow from Model T  to a third universality class --- denote this $U$.  Bl\"ote and Nienhuis suggested that this scenario occurred, and that $U$ should be the ``branch 3'' fixed point for which exact results are available \cite{blote nienhuis, warnaar, blote batchelor nienhuis 98}. This critical point has been revisited very recently by Vernier et al., and shown to have an unusual scaling limit \cite{vernier new look}. In this scenario the presence of incipient Ising order  then gives a natural explanation for why $U$ is different from the generic DS behaviour \cite{blote nienhuis, blote batchelor nienhuis 98, vernier new look}.

Note that we have already ruled out a third  scenario, namely that the flow is from the Bl\"ote Nienhuis fixed point to the fixed point of Model T.

The ISAT multicritical point, which allows crossings and is described by $\rp^{N-1}$ rather than $\cp^{N-1}$ provides a simpler setting for investigating some of the issues of Ising ordering \cite{ising footnote, loops with crossings}.

We emphasise that these questions about Model T, while fascinating, are only indirectly relevant to our basic topic of the  generic collapse behaviour.  From this point of view, the possibility of Ising order in the collapsed phase is a lattice artifact. The true DS fixed point is robust, and the Ising ordering plays no role there. We now return to questions about generic models.

\section{Models with Crossings}
\label{crossings section}

Since in a realistic situation polymers will not be \textit{strictly} confined to 2D, we  expect the chain to be able to cross itself, perhaps at some energy cost (Fig.~\ref{withwithoutcrossings}, right). To understand how this affects the universal behaviour, and also to clarify the relevance of de Gennes' tricritical $\mathrm{O}(n\rightarrow 0)$ model to 2D polymer collapse, we now perturb the square lattice models by allowing crossings. (Note that a crossing is \textit{not} the same as a branching \cite{branching note}: the polymers we consider are always topologically linear.)

Consider either of the two models on the square lattice. The rules for orienting the strands imply that at a four-leg vertex, the two incoming strands are opposite each other and the two outgoing strands are opposite each other (see e.g. Fig.~\ref{2and4legfig}). Therefore, a crossing between two polymer strands (one of colour index $a>0$ and one of colour index $b>0$)  corresponds to a four-leg vertex where the two incoming links are of colour $a$ and the two outgoing links are of colour $b$ (or vice versa). The corresponding perturbation is
\ba\notag
\cros & \equiv - \sum_{a,b=1}^n X_{aabb} \\
 & =- | \vec Z_\perp^T \vec Z_\perp |^2 + 4(N+2)^{-1} |\vec Z_\perp|^2  + \text{const.},
\end{align}
as we can check on the lattice (Appendix~\ref{app perturbations of CPL}). On its own, this operator gives the symmetry breaking
\be
\mathrm{SU}(N)\rightarrow  \mathrm{O}(n) \times \mathrm{U}(1).
\ee
where the $\mathrm{U}(1)$ is $\vec Z_\perp \rightarrow e^{i\theta} \vec Z_\perp$. (This is \textit{not} a gauge transformation, since the phase multiplies only $\zb_\perp$ and not $Z_0$.) If we start with Model T and make a fully generic perturbation (including $\mathcal{A}$, $\mathcal{S}$ and $\mathcal{C}$), then the symmetry is broken down to $\mathrm{O}(n)$ in the original representation. 

This symmetry is what we would originally have expected from de Gennes.  The resulting RG flow away from the DS fixed point, together with the fact that non-crossing models always have a higher symmetry  (despite the subtlety discussed in Sec.~\ref{hidden symmetry section}) indicates that the DS exponents are unlikely to apply to models with crossings. That is, contrary to what is often assumed, we must allow for crossings in order to see the exponents of de Gennes' tricritical $\mathrm{O}(n)$ model.

One point should be clarified. Just as we found in the case without crossings, it is again possible to choose a lattice field theory representation in which we  avoid introducing the operator $\mathcal{S}$. Then, we in fact retain a $ \mathrm{O}(n) \times \mathrm{U}(1)$ symmetry for generic models with crossings. However, we expect that this extra $\mathrm{U}(1)$ can be neglected when considering the generic collapse transition. That is, we expect the latter can be described by a Lagrangian for a real vector that transforms only under $\mathrm{O}(n)$. Symmetries of the Lagrangian are important because they encode information about microscopic constraints on the polymer configurations: here however, the $\mathrm{U}(1)$ does not appear to encode any additional constraints beyond those encoded in $\mathrm{O}(n)$. (Such a $\mathrm{U}(1)$ can always be included in a model of a single polymer `for free'. The current associated with the $\mathrm{U}(1)$ has a simple interpretation. We decorate the polymer with an arrow indicating the direction of $\mathrm{U}(1)$ current flow, using the rule that the polymer's orientation flips whenever it crosses itself. Current is conserved because each crossing has two outgoing and two incoming strands \cite{note about U(1) and orientations}.)

There is a special class of perturbations of Model T which introduces crossings while preserving the equivalence between polymer and background loops.  (All `smart walk' models --- which have the feature that polymer configurations can be regarded as `deterministic walks in  a random environment' --- preserve the equivalence between the polymer and the background loops. This includes the models related to percolation and the collapse point of the interacting self-avoiding trail.) The equivalence is preserved if all the four-leg perturbations have exactly equal strength:
\be
\attr + \stra+ \cros  = - \sum_{a,b=0}^m X_{aabb}.
\ee
The symmetry breaking is then
\be
\mathrm{SU}(N)\rightarrow \mathrm{SO}(N).
\ee
The RG flow then leads to the interacting self-avoiding trail fixed point (ISAT), which is analytically tractable. Unfortunately, it is not the generic $\theta$ point for polymers with crossings. Viewed as a description of a polymer \cite{ISAT fine tuning footnote}, the ISAT fixed point is extremely unstable: it has an infinite number of RG-relevant perturbations that break the symmetry from $\mathrm{SO}(N)$ to the generic $\mathrm{O}(n)$ \cite{loops with crossings}. Signs of this have been seen numerically \cite{bedini trails multicritical}.

Therefore the generic critical exponents for models with crossings remain unknown. A natural model that does not appear fine-tuned has been studied numerically in Ref.~\cite{bedini more generic trails}. It was conjectured in Ref.~\cite{bedini more generic trails} that the critical exponents were those of the DS universality class. This would be surprising in view of the present results. Further numerical results would be valuable.

\section{Operators in the $\cp^{N-1}$ model}
\label{operators section}

Our analysis of perturbations relied on the fact that all the symmetry allowed operators that could appear in the action were components of $Q$ and $X$. In order to confirm this we must now derive some features of the operator content of the sigma model (about which there is currently limited knowledge). We will see that the correspondence with the loop gas implies some surprising things about operators in this field theory. The operators we need to consider in detail are those with the dimension of the four-leg operator, Sec.~\ref{two types of 4 leg operator}, and the marginal operators, Sec.~\ref{marginal operators} (Ref.~\cite{read saleur exact spectra} shows there are no other relevant eigenvalues in the spectrum, apart from $y_2$). We will see that one of the marginal operators  is an interesting parity-odd version of the two-leg operator.

\subsection{Two Types of Four-Leg Operator}
\label{two types of 4 leg operator}

Consider the operators in the field theory which correspond to four-leg operators in the loop model. These are operators whose two-point function gives the probability that $r$ and $r'$ (or rather small regions around $r$ and $r'$) are joined by four strands of loop. At $N=1$, they have scaling dimension $x_4=5/4$.

In the field theory, the obvious operator of this type is $X_{aa'bb'}$ described above: the traceless part of $Z_aZ_{a'}Z^*_bZ^*_{b'}$, which is  invariant under parity. Indeed it is straightforward to check that a lattice  operator with the same symmetries as $X$ allows us to write the 4-leg watermelon correlator in the loop model.  (Strictly speaking the lattice operators cannot have the full symmetry of $X$, since complete invariance under spatial rotations only emerges in the continuum, but this will not be important in what follows.)

\begin{figure}[t]
\includegraphics[width=0.85\linewidth]{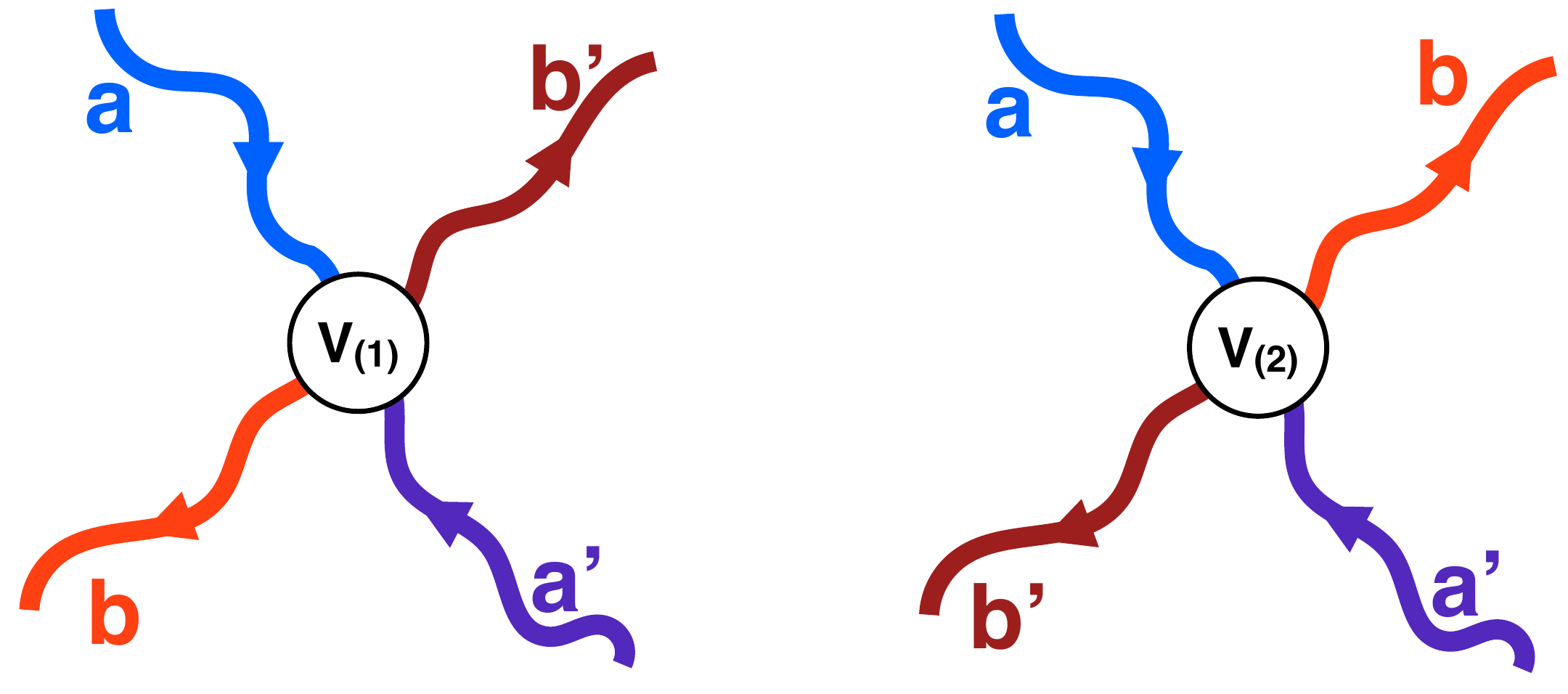}
 \caption{Two types of four leg vertex, distinguished by the ordering of the indices, and related by parity (reflections). The parity-even and parity-odd four leg operators correspond to  the sum and difference respectively, $V_{(1)} \pm V_{(2)}$.}
 \label{V1V2}
\end{figure}

Surprisingly, the sigma model also contains a parity-odd four-leg operator, which we denote $Y$, with the same scaling dimension \cite{springer thesis}. In the field theory, this operator has the following symmetry:
\be
Y_{aa'bb'} = 
i \epsilon_{\mu \nu} \big( Z_a \overset{\leftrightarrow}\nabla_\mu Z_{a'} \big)
\big( Z^*_b \overset{\leftrightarrow}\nabla_\nu Z^*_{b'} \big) -\text{trace terms}.
\ee
Here $A \overset{\leftrightarrow}\nabla_\mu B = A \nabla_\mu B - (\nabla_\mu A) B$.  Unlike $X$, this tensor changes sign under parity, and also under the exchanges $a\leftrightarrow  a'$ and $b\leftrightarrow b'$. It is easily checked to be gauge invariant. Once the trace terms are subtracted, $Y$ forms an irreducible representation of dimension
\be
d_{Y} = \f{N^2 (N+1)(N-3)}{4}.
\ee
From the point of view of field theory it is surprising that this operator, which has completely different symmetry properties and a different number of derivatives, has the same scaling dimension as $X$. This is in fact true for all $N\leq 2$, as we now argue on geometrical grounds. 

Consider a component $X_{aa'bb'}$ of $X$, with all indices distinct. Graphically, this is a vertex with incoming strands of colour $a$ and $a'$, and outgoing strands of colour $b$ and $b'$. Crucially, the no-crossing constraint means that the outgoing strands are opposite each other (Fig.~\ref{V1V2}). This leaves two possibilities for the ordering of the colour indices as we go around the vertex anticlockwise, starting with $a$. Either the colours occur in the order $a,b,a',b'$, or in the order $a,b',a',b$. We may in fact define two distinct operators corresponding to the two orderings, which we denote $V_{(1)}$ and $V_{(2)}$. We may take each to be invariant under spatial rotations, but parity exchanges $V_{(1)}$ and $V_{(2)}$. The operator $X_{aa'bb'}$, which is invariant under parity, is then
\be
X_{aa'bb'} = V_{(1)}+ V_{(2)}.
\ee
But there is also a parity-odd operator,
\be
Y_{aa'bb'} = i(V_{(1)} - V_{(2)}).
\ee 
Note that $Y$ also changes sign under either of the exchanges $a\leftrightarrow a'$, $b\leftrightarrow b'$. These symmetry properties identify it with a component of the operator $Y$ defined above (up to normalisation).

Now consider the correlators $\langle V_{(i)} (r) V^*_{(j)}(r')\rangle$, where the conjugate operators $V_{(i)}^*$ are obtained by reversing the arrows on the strands. These correlators are sums over loop configurations in which the legs of corresponding colour at the two vertices are joined. But the key point is that, because of the no-crossing constraint, no such configurations are possible if $i=j$. We also have $\langle V_{(1)} (r) V^*_{(2)}(r')\rangle=\langle V_{(2)} (r) V^*_{(1)}(r')\rangle$. This implies
\be
\< X_{1234}(r) X_{1234}^*(r')\> = \< Y_{1234}(r) Y_{1234}^*(r')\>.
\ee
Therefore the scaling dimensions of $X$ and $Y$ are equal. This argument holds for any $N$, and generalises immediately to the supersymmetric versions of the sigma models (where we do not have to use the replica-like continuation from $N \geq 4$ to the desired value of $N$).

Note that the argument only shows that the two-point functions of $X$ and $Y$ are the same. More complex correlation functions will reveal the difference between the two operators.

The above argument applies for general $N$. In the special case $N=1$ we may also see that there are additional operators with scaling dimension $x_4$ (i.e. beyond $X$) by an alternative argument. This is because the total multiplicity of each scaling dimension must vanish in the limit $N\rightarrow 1$ on general grounds \cite{cardy logs}. The multiplicity of $X$ is zero in this limit ($d_X = 0$), but we encounter the problem that there is another operator whose dimension $x_W$ coincides with $x_4$ when $N\rightarrow 1$. This is simply the operator in the $\Theta$ term, 
\be
W = i \epsilon_{\mu\nu} \tr Q \nabla_\mu Q \nabla_\nu Q,
\ee
which in the percolation language drives the model off criticality. There must therefore be at least one more multiplet, whose multiplicity cancels that of $W$ as ${N\rightarrow 1}$.  This requirement is filled by $Y$, since $d_Y\rightarrow -1$ as $N\rightarrow 1$. The multiplicities of lattice operators in the spin chain formulation have also been discussed \cite{vasseur quantum hall}, reaching similar conclusions about the cancellation of multiplicities, but without clarifying the geometrical relation between $X$ and $Y$ or the role of parity symmetry.

Refs.~\cite{read saleur enlarged symm alg, read saleur assoc alg} revealed an enlarged symmetry algebra  in completely-packed loop models, related to quantum groups, which is independent of the phase the models are in but depends on the loops not crossing \cite{read saleur enlarged symm alg}. This implies larger degeneracies in the spectrum than expected from $\mathrm{SU}(N)$ alone. This must be the deeper explanation for the above phenomenon. The above argument gives intuitive physical picture for this simple case.

We have found that at $N=1$, there are three types of operators, $X$, $Y$ and $W$, all with scaling dimension $x_4$. How do we know that there are not more? Fortunately  we can use the result  of Read and Saleur \cite{read saleur exact spectra} for the multiplicity of each scaling dimension in the SUSY sigma model. This formula gives the total number of linearly independent operators with dimension $x_4$, without determining their symmetry properties. But if we translate the above operators into the supersymmetric language (we find that  the analogues of $W$ and $Y$ form a parity-odd indecomposable representation) and compute their multiplicities, we can check that the value for the  multiplicity in Ref.~\cite{read saleur exact spectra}  is saturated. This simple calculation is done in  Appendix~\ref{susy operators}.  This shows that there are no other operators with dimension $x_4$, and gives an explicit identification of the supersymmetric operators contributing to the multiplicity formula.

(Another unconventional feature of the $\cp^{N-1}$ model, related to the symmetry discussed in Sec.~\ref{hidden symmetry section} and presumably also a consequence of the extended symmetry of Ref.~\cite{read saleur enlarged symm alg}, is that the operator product expansions are more constrained than would be expected from symmetry. On geometrical grounds it is clear that the OPE of $\mathcal{S}$ with itself cannot generate $\mathcal{C}$, although $\mathrm{SU}(N)$ symmetry would allow this. Equivalently, perturbing the action with $\mathcal{S}$ does not generate $\mathcal{C}$ under RG, consistent with the fact that models with crossings show different universal behaviour to those without.)

\subsection{Marginal Operators in the $\cp^{N-1}$ Model}
\label{marginal operators}

Having pinned down the relevant operators that can perturb the field theory for the polymer, we must also consider whether any marginal perturbations can appear. If present, such perturbations could destabilize the fixed point, or give continuously varying exponents. However we will argue that such perturbations are forbidden by symmetry. This also leads us to an operator which may be independently interesting.

The counting of multiplicities of Ref.~\cite{read saleur exact spectra} is a useful starting point. In the supersymmetric theory, the multiplicity of the scaling dimension $x=2$ indicates that there are two marginal operators, each transforming in the adjoint \cite{read saleur exact spectra}. This  will also to be true in the replica formalism. Let us write these so-far unknown operators as matrices, $A_{ab}$ and $A'_{ab}$. The question boils down to whether they are parity even or parity odd. {If} either operator (say $A$) was parity even, we would  have to worry about the possibility of $A_{00}$ appearing in the action, just as $Q_{00}$ can appear (Sec.~\ref{honeycomb stability section}). However, we argue here that $A$ and $A'$ are parity-odd operators. Therefore spatial symmetry prevents them from appearing in the action. (One of them is also a total derivative in any case.) Our strategy is to exhibit two parity-odd marginal operators, which should therefore be identified with $A$ and $A'$. 

\subsubsection{Parity-Odd Two-Leg Operator Related to Winding Angle}

First, we argue that there is a parity-odd analogue of the two-leg operator, which we denote $Q^\text{odd}_{ab}$, and that correlation functions involving this operator are related to winding angles of the critical curves.

\begin{figure}[t]
\includegraphics[width=0.98\linewidth]{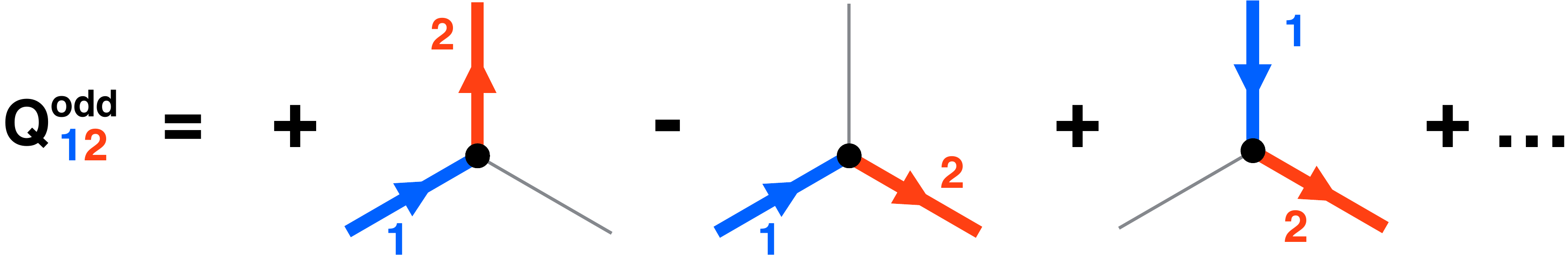}
 \caption{Honeycomb lattice version of $Q^\text{odd}_{12}$. This emits an outgoing leg of colour 2 and absorbs an incoming leg of colour 1, like the 2-leg operator,  but also weights the configuration with a positive or negative sign according to the sign of the turning angle at the node.}
 \label{Qodd}
\end{figure}

Recall that the usual two-leg operator $Q_{ab}$ may be throught of as a vertex with an outgoing $b$ strand and an incoming $a$ strand \cite{diagonal element footnote}. We define $Q^\text{odd}_{ab}$ similarly, except that we weight the vertex by a factor proprtional to the \textit{signed} angle through which the oriented strand turns at the vertex. This does not change the $\mathrm{SU}(N)$ symmetry properties of the operator --- it remains in the adjoint, since it has one fundamental and one antifundamental index. However it becomes manifestly parity-odd, since reflections exchange clockwise and anticlockwise turns. For concreteness, we may take the operator to be defined at vertices of the honeycomb lattice, with left/right turns weighted by $\pm 1$ respectively: see Fig.~\ref{Qodd}. (It is straightforward but not very illuminating to write down such operators in the lattice field theories of Sec.~\ref{lattice partition functions intro}.) In the sigma model, this  operator has the   symmetry of the traceless part of $i\epsilon_{\mu \nu} (Q\nabla_\mu Q \nabla_\nu Q)_{ab}$.
 
We must show that the scaling dimension of this operator, $x_\text{odd}$, is equal to two. To see this, consider the ratio (say on the honeycomb lattice)
\be\label{winding angle correlator lattice}
\mathcal{R} = - \f{\sum_{r, r' (r\neq r')}  \< Q_{12}(0) Q^\text{odd}_{23}(r) Q_{34}(R) Q^\text{odd}_{41}(r') \>}{\< Q_{12}(0)  Q_{21}(R) \>}.
\ee
The correlators in the numerator and deonominator may both be written as sums over  configurations with a loop passing through the sites  at $0$ and $R$. The denominator serves as a partition function for this restricted ensemble. For the correlator in the numerator, one arm of the loop (that from $0\rightarrow R$) also passes through $r$, and the other arm (from $R\rightarrow 0$) passes through $r'$, and the configuration is weighted by the product of the turning angles at these two points. Altogether, $\mathcal{R}$ computes (minus) the expectation value of the product of the total turning angles of the two arms. Up to an $O(1)$ correction which is negligible at large $R$, this is just the square of the winding angle for one of the arms (more precisely, the relevant winding angle is the \textit{sum} of the winding angles about the two points $0$ and $R$). This average is known to scale logarithmically as a result of scale invariance \cite{winding angle 1, winding angle 2}:
\be
\mathcal{R} \sim \< (\text{winding angle})^2\>
\sim \ln R.
\ee
We compare this with the length scaling expected from the scaling dimensions of the operators \cite{contact terms} (c.t. stands for contact terms):
\ba\label{winding angle correlator}\notag
\mathcal{R} & \sim - \f{\int \dd^2 r \dd^2 r'  \< Q_{12}(0) Q^\text{odd}_{23}(r) Q_{34}(R) Q^\text{odd}_{41}(r') \> - \text{c.t.}}{\< Q_{12}(0)  Q_{21}(R) \>} \\ & \sim R^{2(2-x_\text{odd})}.
\end{align}
Comparing with Eq.~\ref{winding angle correlator} indicates that $x_\text{odd}=2$, i.e. that $Q^\text{odd}$ is marginal. Therefore it accounts for one of the two marginal operators sought. 

This argument is not specific to a particular value of $N$ or even to the dense phase. By this reasoning,  \textit{any} conformally-invariant fixed point for non-crossing loops should allow a  parity-odd version of the two-leg operator, with dimension $x_\text{odd} = 2$, in an appropriate field theory representation. This includes for example self-avoiding walks, Ising cluster boundaries etc.

\subsubsection{Effect of Non-Chirality of the Currents}

Next, consider the conserved current $J_{\mu ab}$ associated with global $\mathrm{SU}(N)$ symmetry. Here  $a$, $b$ are $\mathrm{SU}(N)$ indices ($J$ transforms in the adjoint) and $\mu$ is the spatial index. The current has length dimension $- 1$ and satisfies $\nabla.J=0$ as a result of conservation. In a unitary CFT, the current would \textit{also} satisfy $\nabla\times J=0$ \cite{affleck cts symmetries}. In complex coordinates, this leads to $J_z$ being purely holomorphic and $J_{\bar z}$ being purely antiholomorphic. However it is known that this separation into holomorphic and antiholomorphic currents \textit{fails} in the present nonunitary theory \cite{read saleur exact spectra}. Equivalently, $\nabla\times J$ is nonzero as an operator (although it has a vanishing two-point function).

$\nabla\times J$ provides another marginal operator that is manifestly parity odd and transforms in the adjoint. It is distinct from the operator $Q^\text{odd}$ defined above ($Q^\text{odd}$ is not a total derivative, otherwise we could not use it to calculate the winding angle). 

\subsubsection{Implication for Polymers}

We infer that $Q^\text{odd}$ and $\nabla \times J$  correspond to the only marginal scalar operators (that are local in the $\cp^{N-1}$ representation). This saturates the counting of states from Ref.~\cite{read saleur exact spectra}. It follows that there are no marginal perturbations allowed in the action for the polymer problem. This is also what we expect from numerical simulations, which do not see signs of the logarithmic drifts that would be expected for a marginally relevant/irrelevant variable, or the continuously varying exponents that would be expected for an exactly marginal one \cite{caracciolo}.

\section{Outlook}
\label{Conclusions sec}

We have shown that the Duplantier-Saleur exponents for the $\theta$ point are generic for non-crossing polymers, as a result of symmetry enhancement under the RG flow.  This resolves a longstanding question about the stability of the DS point for which  previously there was only numerical evidence \cite{caracciolo}. We have also argued that crossings induce a flow to a new universality class.  Along the way we had to obtain a clearer picture of operators in the $\cp^{N-1}$ sigma model. We also had to resolve some apparent paradoxes about the fine-tuned Model T, which at first sight gives misleading conclusions about the robustness of the DS exponents and about the difference between models with and without crossings. The first of these issues is related to the fact that the same field theory may describe different models, but with a different relationship between polymer and field theory operators in each case. The second issue is related to the fact that the replica-like symmetry of a polymer model can be nontrivially dependent on the choice of field theory mapping.

Many exciting questions remain for the future. Firstly, the full structure of the RG flows for a non-crossing polymer on the square lattice, in the regime where Ising order is playing a role \cite{blote nienhuis}, remains to be understood. Exciting progess has been made very recently on the conformal field theory of the ``branch 3'' fixed point of Bl\"ote and Nienhuis, which appears to be unconventional \cite{vernier new look}. The flow away from Model T may be to this fixed point  \cite{blote nienhuis, vernier new look}: it would be very interesting to have a heuristic understanding of this flow, from the point of view of an effective field theory got by perturbing the $\cp^{N-1}$ Lagrangian.

Another longstanding question concerns certain sequences of multicritical points found in supersymmetric theories, and how to interpret them in terms of polymers \cite{saleur susy, cardy susy}.

Most importantly, models with crossings remain very little understood, despite the fact that a realistic model of a polymer living on a surface or in a quasi-2D geometry will likely include them. Historically such models have been neglected --- perhaps because of the remarkable power of techniques like the Coulomb gas \cite{cg reviews} and Schramm Loewner Evolution \cite{sle reviews}, which only work when crossings are forbidden. The present results motivate further examination of models with crossings. This will be necessary to understand polymer collapse in the fully generic situation, and is likely to reveal novel aspects of 2D criticality \cite{jacobsen read saleur, martins nienhuis rietman, loops with crossings}. 

Finally there are interesting aspects of the $\cp^{N-1}$ field theory and its supersymmetric cousin \cite{read saleur exact spectra} that deserve further study; for example it would be interesting to study the marginal operator $Q^\text{odd}$ introduced here numerically.

After this work was completed, a preprint appeared on dilute loop models \cite{new vernier} --- this addresses different questions to the present paper, but also considers a deformation of the lattice field theory for completely packed loop models \cite{cpn loops short, loops with crossings, deformation footnote}. Also, a pair of numerical studies of the phase diagrams of generalised square \cite{new prellberg 1} and honeycomb lattice models \cite{new prellberg 2} appeared. The results appear consistent with expectations from our analysis. The phase structure found in Ref.~\cite{new prellberg 1} seems to suggest that Scenario II in our Sec.V D is more likely than Scenario I \cite{adjacency of ising ordered phase}.

\acknowledgements

I am very grateful to E. Bettelheim,  J. Chalker and J. Cardy for helpful discussions and comments, and encouragement to finally get this written up, and I also thank S. Vijay for useful discussions.  I acknowledge the support of a fellowship from the Gordon and Betty Moore Foundation under the EPiQS initiative (grant no. GBMF4303).

\appendix

\section{Generic Perturbations of the Honeycomb Model}
\label{honeycomb lattice perturbations}

We begin with the lattice field theory in Eq.~\ref{lattice gauge theory}, which maps to the polymer model $Z_\text{polymer}$ in Eq.~\ref{honeycomb polymer partition function}. We discuss how deforming the Boltzmann weight for the lattice field theory leads to a modified polymer model. First consider simply inserting a factor $y$ as follows:
\ba\label{honeycomb perturbed}
Z(y) = \Tr \bigg\{ \hspace{0mm} &\prod_\text{hexagons H} \hspace{-1mm} \bigg( 1 + \prod_{\<ij\> \in \text{H}} U_{ij} 
\bigg)\times
\\\notag
&\prod_{\<ij\>} \lf 
1 + U_{ij} \lf Z_{0i}^* Z_{0j}^{\phantom{*}} + y \, \zb_{\perp i}^\dag \zb_{\perp j}^{\phantom{\dag}} \ri+ \text{c.c.}
\ri \bigg\}.
\end{align}
In the graphical expansion, each segment of polymer loop ($\zb_\perp$ worldline) now acquires a factor of $y$. We therefore obtain a polymer  with a modified weight per unit length:
\be\label{honeycomb polymer partition function modified}
Z_\text{polymer}(y) =  \sum_{\substack{\text{polymer}\\ \text{configs}}}  2^{-H} y^\text{length}.
\ee
$H$ is the number of hexagons visited by the polymer.  Making $y$ smaller than one takes the model off criticality, so that the polymer becomes of a finite typical size. Taking $y>1$ will drive the model into the dense polymer (i.e. space-filling) phase, where the polymer's length scales with the total area of the lattice.

Varying $y$ is a rather trivial perturbation to the Boltzmann weight. However (\ref{honeycomb perturbed}) illustrates the basic point --- changing the polymer interactions induces \textit{local} interactions in the lattice field theory, which break the symmetry from $\mathrm{SU}(N)$ to $\mathrm{U}(n)$. Next, we must check that \textit{any} local perturbation to the polymer Boltzmann weight maps to a local perturbation in Eq.~\ref{honeycomb perturbed}. Let $n_{\ell}$ be the occupation number of the link $\ell$ in a given polymer configuration: i.e. $n_{\ell}=1$ if the polymer passes through the link and $n_\ell=0$ otherwise. The general perturbed partition function is:
\ba\label{general honeycomb partition function}
Z_\text{polymer}(y, J, K, \ldots) = 
\sum_{\substack{\text{polymer}\\ \text{configs}}}  2^{-H}
\exp A
\end{align}
with
\ba\notag
A =   (\ln y) \sum_\ell n_\ell
&+
\sum_{\ell, \ell'} J_{\ell, \ell'} n_{\ell} n_{\ell'} \\ \notag
&+ 
\sum_{\ell, \ell', \ell''} K_{\ell, \ell', \ell''} n_{\ell} n_{\ell'} n_{\ell''} + \ldots.
\end{align}
Expanding the exponential in these couplings gives a sum of terms proportional to $n_{\ell_1}\ldots n_{\ell_k}$, where all the links can be taken distinct (since $n_{\ell}^2=n_{\ell}$). Therefore we must check that an insertion of $n_\ell$ corresponds to a local operator in the lattice field theory. This follows from the correspondence
\be
n_{ij}  \,\, \longrightarrow \, \,
\f{
U_{ij} \lf \zb_{\perp i}^\dag \zb_{\perp j}^{\phantom{\dag}} + \text{c.c.} \ri
}{
1 + U_{ij} \lf \zb_{i}^\dag \zb_{j}^{\phantom{\dag}} + \text{c.c.} \ri
},
\ee
which we may check by repeating the graphical expansion in the presence of $n_{ij}$ insertions. 

A crude effective action may be obtained by coarse graining Eq.~\ref{honeycomb perturbed} or its perturbed version. We write $U_{ij} = e^{i A_{ij}}$, and expand the logarithm of the Boltzmann weight in $A$, in derivatives of $\zb$, and in the size of the perturbation (see e.g. Refs.~\cite{vortex lines, loops with crossings}). (For a crude picture of the perturbation terms, we may take  $U$ and $\zb$ as spatially constant: then the above formula is simply $n_{ij} \propto |\zb_\perp|^2$, so that for example $\exp  \sum_{ \ell, \ell' } J_{\ell,\ell'} n_{\ell} n_{\ell'}$ generates a quartic potential for $\zb_\perp$ at leading order in $J$.)  

However for our purposes all we need are the relevant operators which appear in the coarse grained action, not the numerical values of the couplings. These operators are determined by symmetry and are given in Sec.~\ref{honeycomb stability section}.

Note that above we have not changed the allowed configurations for the polymer. Allowing configurations in which the polymer crosses itself (on the honeycomb lattice this can happen if for example we allow double occupancy of a link) introduces another relevant perturbation, see Sec.~\ref{crossings section}.

\section{Perturbations of the L-Lattice Model}
\label{app perturbations of CPL}

For the square lattice model of Eq.~\ref{CPL polymer partition function} we will discuss a few illustrative deformations of the Boltzmann weight. Consider first the slightly generalised model
\be\label{CPL polymer partition function modified}
Z_\text{polymer} =  \sum_{\substack{\text{polymer}\\ \text{configs}}} 
A^\text{length} B^\text{no. twice-visited nodes}.
\ee
This  becomes Model T  when $A=1/2$, $B=2$. To obtain this from the lattice magnet in Eq.~\ref{lattice cpn-1 model}, 
\ba \notag
Z_\text{CPL}& = \Tr \prod_\text{nodes} e^{-S_\text{node}}, \\
e^{-S_\text{node}} &=
  (\zb_o^\dag \zb_i) (\zb_{o'}^\dag \zb_{i'}) +  (\zb_{o'}^\dag \zb_i) (\zb_{o}^\dag \zb_{i'}),
\end{align}
we note that each term in $e^{-S_\text{node}}$ of the form
\be
W(\zb_i, \zb_{i'}, \zb_o, \zb_{o'})  \equiv (\zb_o^\dag \zb_i) (\zb_{o'}^\dag \zb_{i'}) 
\ee
 can be expanded into terms which, depending on the values of the indices in the inner products, correspond either to (1) two segments of polymer passing through the node, or (2) one segment of polymer and one segment of background loop, or (3) two segments of background loop. The weights of these possibilities can be adjusted by replacing the above with \cite{loops with crossings}
\ba\notag
 W(&\zb_i, \zb_{i'}, \zb_o, \zb_{o'})  \, \longrightarrow \, &\\ \notag
&2A^2  B(\vec Z_{o\perp}^\dag \vec Z_{i\perp})(\vec Z_{o'\perp}^\dag \vec Z_{i'\perp}) +2A (Z_{o0}^*Z_{i0})(\vec Z_{o'\perp}^\dag \vec Z_{i'\perp}) \\
+&2A(\vec Z_{o\perp}^\dag \vec Z_{i\perp})(Z_{o'0}^*Z_{i'0}) +(Z_{o0}^*Z_{i0})(Z_{o'0}^*Z_{i'0}). 
\end{align}
This is analogous to the perturbations discussed in App.~\ref{honeycomb lattice perturbations}, and breaks the symmetry down to $\mathrm{U}(n)$. The graphical expansion goes through straightforwardly and gives the desired modification to the polymer Boltzmann weight.

Next consider the introduction of straight segments for the polymer. This is achieved by the modification
\ba\label{AB modification}
&e^{-S_\text{node} }\,\longrightarrow \, 
e^{-S_\text{node} }
  +  C \, R(\zb_i, \zb_{i'}, \zb_o, \zb_{o'}), 
\end{align}
where $C$ is some weight and 
\ba\notag
R( \zb_i,& \zb_{i'},  \zb_o, \zb_{o'})  \\
=& \, (Z_{o0}^* Z_{o'0}^*) (\vec Z_{i\perp}^T \vec Z_{i'\perp}) 
+ (\vec Z_{o\perp}^\dag \vec Z_{o'\perp}^*) ( Z_{i0}^T  Z_{i'0}). 
\end{align}
Note that this is a lattice analogue of Eq.~\ref{Ostraight}. This modification preserves the  gauge invariance of the total Boltzmann weight. The colour index of a strand is still preserved along its length, but now the two outgoing links lie on one strand, and the two incoming links lie on a different strand --- the polymer and background segments cross at the node. The pattern of complex conjugation means that unitary symmetry is broken.

Finally, we allow nodes where the polymer crosses itself. This corresponds to adding to the node term a multiple of the expression
\be
 (\vec Z_{o\perp}^\dag \vec Z_{o'\perp}^*)(\vec Z_{i\perp}^T \vec Z_{i'\perp}).
\ee
The corresponding additions to the continuum action follow on symmetry grounds. For a crude estimate of the couplings in the perturbed sigma model, we can evaluate the Boltzmann weight with $\zb$ spatially constant, and we see that the terms discussed in Secs.~\ref{section on additional perturbation in square lattice model},~\ref{crossings section} appear in the action (i.e. the logarithm of the Boltzmann weight) with the expected signs.

\section{More Details on Second Square Lattice Model}
\label{appendix square perc model}

\subsubsection{Boltzmann Weight for Associated Polymer Model}

At $N=1$, the model in Eq.~\ref{second square lattice model} maps to a model for a polymer ring. Any loop drawn on the square lattice corresponds to an allowed polymer conformation so long as no edge is visited more than once, no site is visited more than twice, and the loop does not cross itself. The Boltzmann weight for a given polymer configuration is simply the probability of that loop appearing in the loop gas. This is easily evaluated using the mapping to a percolation problem:
\be
Z_\text{LP} = \sum_{\substack{\text{polymer}\\ \text{configs}}} 
\lf \f{1}{2} \ri^{N_F} 
\lf \f{1}{2} \ri^{N_T}
\prod_{k=1}^4 \lf \f{1}{2} + \f{1}{2^k} \ri^{N_k}. 
\ee
Here, $N_F$ is the number of faces of the square lattice that contain at least one edge visited by the polymer. $N_T$ is the number of sites that are visited twice by the polymer. $N_k$ is the number of faces which contain $k$ sites visited by the polymer but no link visited by the polymer. This Boltzmann weight is somewhat cumbersome to write down (it has a simpler representation, described below) but it is perfectly local, and the model has the advantage that --- thanks to the mapping to percolation --- we know that the statistics of the polymer ring are those of the DS point. 

\subsubsection{Lattice Gauge Theory}

The loop gas in Eq.~\ref{second square lattice model} maps to a lattice gauge theory  identical to Eq.~\ref{lattice gauge theory}, modulo the substitution of the square for the honeycomb lattice (and square faces for hexagons). 

\begin{figure}[t]
 \begin{center}
       \includegraphics[width=0.8\linewidth]{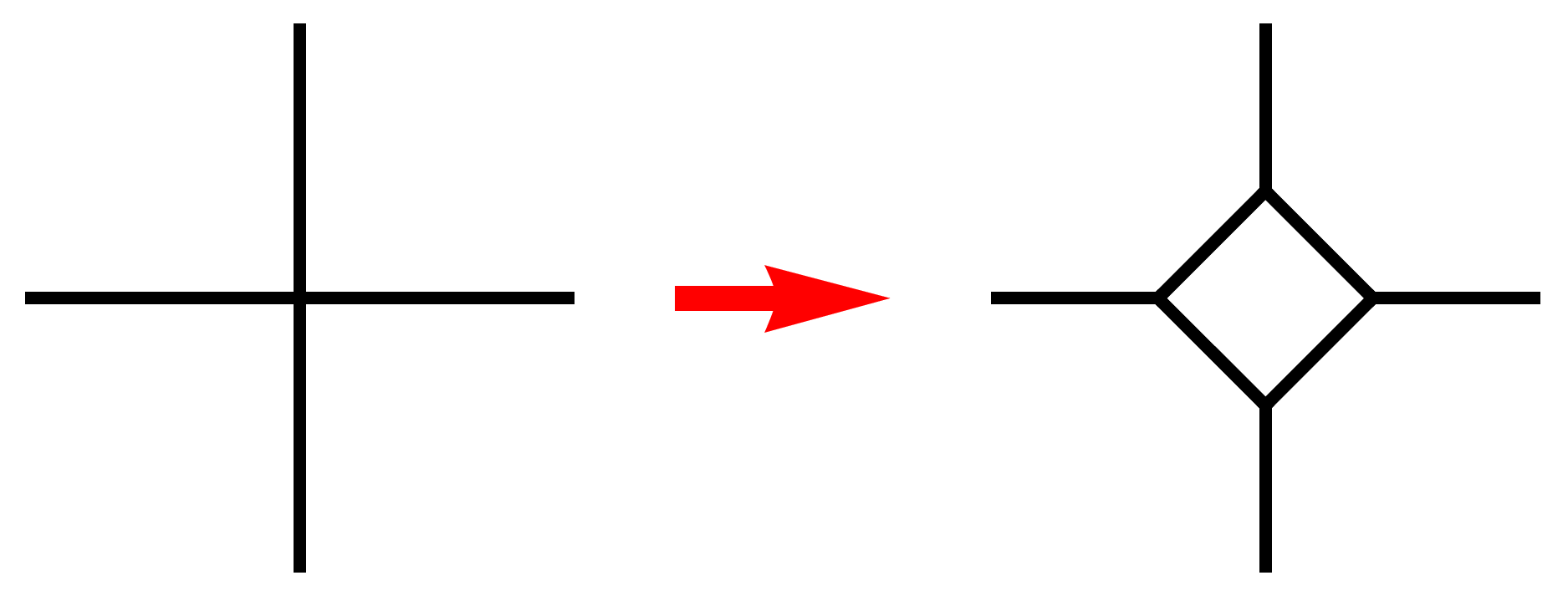}
 \end{center}
 \caption{It is convenient to expand the nodes of the square lattice so that the polymer becomes strictly self-avoiding not only on the links but also on the nodes.}
 \label{resolved node}
\end{figure}

In order to consider the most general perturbations of the Boltzmann weight, however, it is convenient to map the loop gas to field theory in a slightly different way. First, we resolve the nodes as in Fig.~\ref{resolved node}, inserting a small diamond at each vertex so that the lattice becomes three-coordinated. Now we consider a straightforward percolation model in which we randomly colour (all) the faces of this new lattice black or white with equal probability. This percolation problem maps to the previous one in a trivial way. Previously, there were two possibilities for how to connect up the clusters when two of them met at a corner; now these two possibilities correspond to the two colourings of the diamond. Similarly, the loop gases are simply related at $N=1$. The polymer partition function also takes a simple form if we regard the polymer as living on the new lattice --- the weight is simply $(1/2)$ raised to the power of the number of faces visited. We emphasise that this is simply a different and more convenient representation of the \textit{same} polymer model that we started with.

The loop gas on the new lattice again has a lattice gauge theory representation like Eq.~\ref{second square lattice model}:
\be
Z = \Tr \prod_\text{faces F} \hspace{-1mm} \bigg( 1 + \prod_{\<ij\> \in \text{F}} U_{ij} 
\bigg)
\prod_{\<ij\>} \lf 
1 + U_{ij} \zb^\dag_i \zb_j + \text{c.c.}
\ri.
\ee
The product over faces now runs over both  4-sided and 8-sided faces.

The reason for adopting this representation is that  the polymer is now strictly self-avoiding (on sites as well as on links). This means that the configuration is completely determined by which links are occupied, which was not the case on the original square lattice (we needed to specify both which links were occupied and also how the strands were connected up at twice-visited nodes). This means that we can use the mapping between operators described in Sec.~\ref{honeycomb lattice perturbations} to map any local perturbation of the polymer Boltzmann weight to a local perturbation in the lattice field theory.

\section{Note on Operators with Dimension $x_4$ in SUSY Language}
\label{susy operators}

In the text we discussed three operators with dimension $x_4$, namely the parity-even 4 leg operator $X_{aa'bb'}$, the parity-odd four leg operator $Y_{aa'bb'}$, and the parity-odd singlet operator in the $\Theta$ term, $W$. The total multiplicity of these operators was $d_X + d_Y + 1$, which tends to zero in the limit $N\rightarrow 0$. This vanishing of the total multiplicity is required by general constraints on the spectrum of theories with central charge zero \cite{cardy logs}. (In fact the multiplicities should  vanish separately in the parity-odd and parity-even sectors, $d_X\rightarrow 0$ and $d_Y + 1\rightarrow 0$.) This strongly suggests that $X$, $Y$ and $W$ form the full set of operators with dimension $x_4$. To make absolutely sure, we translate these statements into the language of the SUSY sigma model, where we can use the results of Ref.~\cite{read saleur exact spectra} for the counting of states. 

In the SUSY model,  the $\cp^{n_b-1|n_f}$ sigma model, the spin $\zb$ is upgraded to a superspin 
$\Psi = (Z_1, \ldots Z_{n_b}, \chi_{1}, \ldots, \chi_{n_f})$ with $n_b$ bosonic and $n_f$ fermionic components, which transforms under the superunitary group; see Ref.~\cite{read saleur exact spectra} for more information. The SUSY sigma model describes the same physics as the replica sigma model so long as $n_b-n_f  = N$. The value of $n_b+n_f$ is arbitrary; increasing $n_b+n_f$ gives a richer spectrum of operators but does not change the partition function or the mapping to the loop gas. The matrix $Q_{ab}$ becomes $\Psi_a \Psi_b^\dag - N^{-1} \delta_{ab}$, which is supertraceless.  (To form  the supertrace, indices are contracted up not with $\delta_{ab}$ but with $\eta_{ab} = \eta_a \delta_{ab}$, where $\eta_a$ is $+1$ for bosonic and $-1$ for fermionic values of the index.) At $N=1$, the multiplicity of the scaling dimension $x_4$ is \cite{read saleur exact spectra}
\be\label{d4 susy}
d_{x_4} = 4 n_f (2n_f^3 + 4 n_f^2 + n_f - 1).
\ee

We now translate our operators $X$, $Y$ and $W$ into the SUSY language, calculate their total multiplicity, and confirm that at $N=1$ it saturates $d_{x_4}$. This shows that there cannot be any other operators with dimension $x_4$. (In the SUSY representation, all multiplicities are of course positive, since there is no replica-like analytic continuation.) 

First, consider the matrix $M_{aa'} = \Psi_a\Psi_{a'}$. This is symmetric in $a$, $a'$ unless both of the indices are fermionic, in which case it is antisymmetric. Let us call this a ``symmetric'' tensor in quotation marks, and a tensor which is antisymmetric except when both indices are fermionic an ``antisymmetric'' tensor. The number of independent components in $M_{aa'}$ is therefore
\be
n_b(n_b+1)/2 + n_b n_f + n_f(n_f-1)/2.
\ee
Next, consider $\widetilde M_{ab} = \Psi_a \overset{\leftrightarrow}\nabla_\mu \Psi_b$ (the spatial index will not play a role). This object is ``antisymmetric''. The number of components is got by exchanging $n_b\leftrightarrow n_f$ in the above formula.

The SUSY version of $X$ is given by taking $\widetilde X = \Psi_a\Psi_{a'} \Psi^\dag_{b}\Psi^\dag_{b'}$ and subtracting appropriate terms to make it supertraceless (meaning that it vanishes when an index on one of the $\Psi$s is contracted with an index on one of the $\Psi^\dag$s) and therefore irreducible under the superunitary symmetry. Here such subtractions are possible for positive $N$, but see below.

The number of independent components in $\widetilde X$ is simply the square of the number of components in $M_{aa'}$ above. However in making it traceless we remove $(n_b+n_f)^2$ components. This is simply the number of independent components that are left when we contract one pair of indices on $\widetilde X$ (all of these are  set to zero when we make $X$ supertraceless). Therefore, when $N=n_b-n_f = 1$, the dimension of the  irreducible representation in which $X$ transforms is
\be\label{susy dx}
d_X = 4 n_f^2 (n_f+1)^2.
\ee

Next consider $\widetilde Y_{aa'bb'} = \epsilon_{\mu \nu} (\Psi_a \overset{\leftrightarrow}\nabla_\mu \Psi_{a'}) (\Psi^\dag_b \overset{\leftrightarrow}\nabla_\nu \Psi^\dag_{b'})$. The number of independent components is the square of the number in $\widetilde M$. When we  perform subtractions to make $\widetilde Y$ supertraceless, naively we again remove $(n_b+n_f)^2$ components. This is correct for $N>1$: in that case $\widetilde Y$ splits into an irreducible fully traceless object (the analogue of $Y$ in the replica theory), a singlet (the analogue of $W$) and a two-index (adjoint) object which is not of interest to us here. For $N=1$ however the invariant four-index tensor with the appropriate ``antisymmetry'' properties --- denote it $c_{aa'bb'}$ --- has vanishing trace, so subtracting $c_{aa'bb'}W$ does not change the trace of $\widetilde Y$. As a result, we expect that the resulting the four-index object forms an indecomposable representation which includes the singlet $W$. (For a simpler analogue, consider the matrix $Q_{ab} = \Psi_a\Psi^\dag_b - (n_b - n_f)^{-1} \delta_{ab} \Psi^\dag \Psi$. When $n_b>n_f$, the subtraction ensures that the supertrace of $Q$ vanishes. However when $n_b=n_f$, the supertrace of the identity vanishes, so we cannot make $Q$ traceless. This means that at $N=0$, the supermatrix $\Psi \Psi^\dag$ forms a single indecomposable representation of dimension $(n_b+n_f)^2$ which includes the singlet $\Psi^\dag \Psi$.)  The dimension of this representation is ($N=1)$
\be\label{susy dyw}
d_{Y/W} = 4 n_f (n_f +1) (n_f^2 + n_f -1).
\ee
The notation indicates that this indecomposable representation in the SUSY model subsumes the analogues of \textit{both} $Y_{aa'bb'}$ and $W$ in the replica formulation. (Recall that $Y$ was the parity odd four-leg operator and $W$ was the singlet which appears in the $\Theta$ term.)

From the analysis in the main text, we know that the operators discussed above all have scaling dimension $x_4$. The scaling dimensions in the replica theory and the SUSY model are of course the same. (For example this follows from the fact that they are related to the same correlators in the loop gas.) Adding up the multiplicities $d_X$ and $d_{Y/W}$ gives perfect agreement with Eq.~\ref{d4 susy} for $d_{x_4}$. This confirms that we have identified the complete set of operators with this scaling dimension, and also explains where Eq.~\ref{d4 susy} comes from physically.


\begin{thebibliography}{250}
\bibitem{theta pi footnote} This is associated with the fact that adjacent loops in Fig.~\ref{looppolcorrespondencehoneycomb}, left, have opposite orientations --- or in the 1+1D language, the fact that particles and antiparticles alternate in space \cite{shankar murthy}.
\bibitem{theta term footnote} $\Theta$ changes sign under reflections, but since $\Theta$ is equivalent to $\Theta +2\pi$ (up to a change of boundary conditions), $\Theta = \pi$ is compatible with spatial symmetry. A trick to obtain the value of $\Theta$ is as follows. Take a system with the topology of the disk (without dangling links at the boundary). Then, if $Q$ is smoothly varying, it is possible to find a gauge where $\zb$ is smoothly varying, and furthermore where the Boltzmann weight is real and positive for all the nodes in the interior. The imaginary part of the action then comes only from the Boltzmann weight for the boundary links. Labelling them by $l$, this is $\prod_{l} \zb^{\dag}_{l+1} \zb_l$. in the continuum this becomes an exponential of the line integral of $\zb^\dag \nabla \zb$ along the boundary. This is precisely the $\Theta$ term in the action, which is a total derivative when written in terms of $\zb$ (so reduces to a boundary integral if we can choose a gauge where $\zb$ is smoothly varying). This term is pure imaginary. See Refs.~\cite{deconfined paper,springer thesis} for a discussion of similar issues in a  3D loop model.
\bibitem{operator identification footnote} The true scaling operators of the fixed point theory are not of course identical with simple expressions in the sigma model fields, as different operators with the same symmetry properties can mix (for example $Q_{ab}$ could mix with $Q_{ab}\tr (\nabla Q)^2$, etc.). However this is not important for the symmetry analysis.
\bibitem{rotational invariance} An operator defined on a link is not of course rotationally invariant: the leading contribution to this lattice operator is the rotationally invariant continuum operator which we denote $Q$.
\bibitem{bead spring footnote} One could even consider off-lattice (e.g. bead-and-spring type) models. While the regularization of the appropriate gauge theories would be  tedious to perform explicitly, it is clear that the same symmetry arguments would apply.
\bibitem{ising domain wall footnote}  In the field theory obtained from the completely-packed loop representation (but not in that obtained from the incompletely packed representation!) a domain wall in the Ising order parameter is essentially a worldline of $Z_0$ \cite{loops with crossings}, so the the presence or absence of Ising order in the {dense} phase is determined by whether this component is massive or condensed.
\bibitem{integrate out footnote} This is easiest to see in the formulation of the $\cp^{N-1}$ sigma model that includes an explicit fluctuating gauge field.
\bibitem{twist footnote} We will discuss this in more detail elsewhere.
\bibitem{ising footnote} When the polymer can cross itself, the definition given in the text for the Ising order parameter  needs to be slightly modified: we take the orientation given to the polymer to flip every time it crosses itself.
\bibitem{branching note} If included, 4-leg branchings would induce the perturbation  $\sum_{a>1} X_{aaaa}$, breaking the part of the symmetry group which mixes different components to permutations, $S_n$.
\bibitem{note about U(1) and orientations} This rule for orienting the polymer is also that which is necessary  in order to extend the definition of $\sigma$ in  Sec.~\ref{discussion of fate of model T} to the case with crossings \cite{ising footnote}. That is, the sign of $\sigma$ is determined by whether the $\mathrm{U}(1)$ current is parallel or antiparallel to the orientation of the link in Fig.~\ref{looppolcorrespondence}.
\bibitem{ISAT fine tuning footnote} While the interacting self-avoiding trail is fine-tuned when regarded as a model for a polymer, it can also be regarded as a model for a deterministic walk in a random environment --- i.e. a Lorentz lattice gas --- and in this interpretation it is {not} fine tuned.
\bibitem{contact terms} Naively coarse-graining Eq.~\ref{winding angle correlator lattice} we might expect a contact term, with a nonuniversal coefficient, scaling like $\<Q(0)Q(R)\>^{-1} \int \dd^2 r \< Q(0) X(r) Q(R)\>$. But this would give a result of order $\mathcal{R} \sim R^{y_4} = R^{3/4}$ (which is the scaling of the number of close approaches of the two arms). Since this is much larger than the correct result, namely $\ln R$, the coefficient of this term must be zero when the lattice expression is correctly coarse-grained. Equivalently, the correct continuum limit should be $\lim_{\epsilon\rightarrow 0} (\int_{|r-r'|>\epsilon} \dd^2 r \dd^2 r' \< Q_{12}(0) Q^\text{odd}_{23}(r) Q_{34}(R) Q^\text{odd}_{23}(r') \> - (8\pi c/3 \epsilon^{3/4}) \int \dd^2 r \< Q_{12}(0) X_{2431}(r) Q_{34}(R)\>)$, where $c$ is the appropriate OPE coefficient.
\bibitem{adjacency of ising ordered phase} Since the true DS transition is into the collapsed phase without Ising order, scenario (I) in Sec.~\ref{discussion of fate of model T} would imply that the two types of collapsed phase, $\<\sigma\>=0$ and $\<\sigma\> \neq 0$, both existed arbitrarily close to Model T --- i.e. that Model T represented the point where the Ising phase transition line between them terminated.
\end{thebibliography}
\end{document}